\documentclass[runningheads]{llncs}

\usepackage{hyperref}

\usepackage{geometry}
\usepackage{color}

\usepackage{tabularx}
\newcolumntype{D}{>{\centering\arraybackslash}m{2.1cm}}

\newcolumntype{U}{>{\centering\arraybackslash}m{0.5cm}}
\newcolumntype{S}{>{\centering\arraybackslash}m{1 cm}}
\newcolumntype{M}{>{\centering\arraybackslash}m{1.2cm}}
\newcolumntype{P}{>{\centering\arraybackslash}m{6cm}}
\newcolumntype{C}{>{\centering\arraybackslash}m{0.4cm}}

\usepackage[labelfont={bf}]{caption}

\usepackage{breqn}

\usepackage[ruled,vlined]{algorithm2e}

\let\oldnl\nl
\newcommand{\nonl}{\renewcommand{\nl}{\let\nl\oldnl}}

\newlength\mylen
\newcommand\myinput[1]{%
  \nonl\settowidth\mylen{\KwIn{}}%
  \setlength\hangindent{\mylen}%
  \hspace*{\mylen}#1\\}
  
\SetKw{KwBy}{by}

\graphicspath{{./figures/}}

\makeatletter
\def\ps@pprintTitle{%
  \let\@oddhead\@empty
  \let\@evenhead\@empty
  \def\@oddfoot{\reset@font\hfil\thepage\hfil}
  \let\@evenfoot\@oddfoot
    }
\makeatother

\begin{document}
\title{Auto-tune POIs: Estimation of distribution algorithms for efficient side-channel analysis}
%
%
\author{Unai Rioja\inst{1,2} \and
Lejla Batina\inst{1} \and
Jose Luis Flores\inst{2} \and
Igor Armendariz\inst{2}}
\authorrunning{Rioja et al.}
%
\institute{Digital Security Group, Radboud University, Nijmegen, The Netherlands \email{\{unai.riojasabando,lejla.batina\}@ru.nl} \and
Ikerlan Technological Research Centre, Arrasate-Mondragón, Gipuzkoa, Spain \email{\{urioja,jlflores,iarmendariz\}@ikerlan.es}}
\maketitle              
\begin{abstract}
Due to the constant increase and versatility of IoT devices that should keep sensitive information private, Side-Channel Analysis (SCA) attacks on embedded devices are gaining visibility in the industrial field. The integration and validation of countermeasures against SCA can be an expensive and cumbersome process, especially for the less experienced ones, and current certification procedures require to attack the devices under test using multiple SCA techniques and attack vectors, often implying a high degree of complexity.

The goal of this paper is to ease one of the most crucial and tedious steps of profiling attacks i.e. the points of interest (POI) selection and hence assist the SCA evaluation process. To this end, we introduce the usage of Estimation of Distribution Algorithms (EDAs) in the SCA field in order to automatically tune the point of interest selection. 
We showcase our approach on several experimental use cases, including attacks on unprotected and protected AES implementations over distinct copies of the same device, dismissing in this way the portability issue.

\keywords{SCA \and Profiling Attacks \and Template attacks \and Portability \and POI selection \and EDAs \and AES \and Masking countermeasure}
\end{abstract}
\section{Introduction}

Today, with the growing presence of Industry 4.0 and IoT devices in our lives, more and more customers and product developers are faced with the utmost importance of the cybersecurity of small embedded devices. Consequently, this has led to a general increase in the interest in Side-Channel Analysis (SCA) and countermeasures, and in general in all physical attacks on among others, IoT devices. The main challenge is in the integration and validation of countermeasures against SCA being a long, expensive, and complex process in practice. Assessing the security of an electronic system against SCA is a problem that requires various skills and expertise from very different fields (electronics and hardware, signal processing, statistics, cryptography, deep learning, etc.) and in which the experience of a security analysis professional is often a crucial part of this operation. 

This is partly due to current certification processes like EMVCo\cite{EMVCo} or Common Criteria (CC)\cite{CC} requiring to evaluate the robustness of the \textit{Device Under Test} (DUT) by performing a battery of distinct side-channel attacks (such as differential power analysis (DPA)~\cite{kocher1999dpa}, correlation power analysis (CPA)~\cite{brier2004cpa}, mutual information analysis (MIA)~\cite{Gierlichs2007mutual,lejla2011mia}, template attacks (TAs)~\cite{Chari2002template,Choundary2018efficient}, deep learning-based attacks (DL-SCA)~\cite{Maghrebi2016BreakingCI,picek2018ontheperformance,masure2019comprehensive}). The motivation is to quantify the security of a system against SCA by taking into account whether the attacks are successful or not and the number of resources that they require. This approach is overseen by organizations like ANSSI \cite{ANSSI} and BSI \cite{BSI} and the amounts of time and resources needed for performing this kind of evaluations is constantly growing (with the new attacks and techniques being proposed), making infeasible to assess the security of a system against SCA in a fast, efficient and low-cost manner. Trying to mitigate this issue, several leakage assessment techniques (like Test Vector Leakage Assessment, TVLA~\cite{Gilbert2011tvla}) have arisen intending to assess whether a device leaks information through side channels or not. The problem is that the security of a system against SCA cannot be assessed simply by applying these techniques: leakage assessment tests only determine whether there is any leakage in the power traces, without specifying whether that specific leakage is critical or giving any hints on how to do exploit it. Therefore, leakage assessment tests do not solve the problem since to evaluate the security of an embedded system against SCA properly it is mandatory to attack it exhaustively with known SCA techniques (including TAs), with the complexity that this entails. 

The biggest challenge for TA in particular, but also for PA in general, is the finding of proper time samples containing the leakage information (usually named Points Of Interest, POI). 
Thus, to ease the evaluation process in general, and TAs specifically, we propose to perform POIs tuning together with the template building and key recovering steps, automatically.
This allows expert evaluators to save time and parallelize tasks (improving the efficiency of the process) but also helps technicians without a deep knowledge of all the basics involved in these methods to implement TAs properly.

Thus, the main contribution of this paper is in introducing a novel advanced and automatized search strategy for the Point of Interest (POI) tuning issue. We demonstrate that this approach straightforwardly provides state-of-the-art results by searching the best groupings of POIs in the space shaped for all possible groupings.
More specifically, the entire contribution of this work can be divided into the following parts:

\begin{enumerate}
    \item We propose a novel approach for the Point of Interest (POI) tuning issue, which is fully automated. Thus, this approach not only improves the state-of-the-art in terms of performance but also mitigates complexity issues. This is accomplished by applying Estimation of Distribution Algorithms (EDAs)~\cite{Muhlenbein1996edas,Larranaga2002edas,Pelikan2003hierarchical,Lozano2006edas,Coffin2008eda} which, to the best of our knowledge, have never before been applied in the SCA field. We name this technique Estimation of Distribution Algorithm-Profiling Attack (EDA-Based PA). 
    
    \item We demonstrate our approach on different datasets and devices, including unprotected hardware AES implementation on FPGA (AES\_HD data set\footnote{AES\_HD Dataset: \url{https://github.com/AESHD/AES_HD_Dataset}}) and protected (masked) software AES implementation on microcontroller (ASCAD data set\footnote{ASCAD Dataset: \url{https://github.com/ANSSI-FR/ASCAD}}~\cite{prouff2018ascad}). Thus, we prove that our EDA-Based PA is suitable for different implementations and leakage models.
    
    \item Moreover, in order to demonstrate the applicability of our method in a more realistic context, i.e. considering portability (see Sec.~\ref{subsec.Portability}), we introduce a new dataset: the \textit{AES\_PT} dataset\footnote{AES\_PT Dataset: \url{https://github.com/urioja/AESPT}}(Sec. \ref{introAESPT}). \textit{AES\_PT} is the first open dataset for SCA which includes power analysis traces of four different copies of the same hardware device: a high-performance ARM\textsuperscript{\tiny\textregistered} Cortex\textsuperscript{\tiny\textregistered}-M4 32-bit RISC microcontroller (STM32F411VE~\cite{UCdatasheet}). This dataset was created with the idea of making ``realistic'' TAs and therefore includes subsets of traces of each clone device performing unprotected and protected AES-128 implementations, with both fixed and random cryptographic keys. 
    
    \item Besides, we also demonstrate the suitability of our technique in this \textit{AES\_PT} dataset, and hence in a portable template attack scenario. Therefore, we show how even in this real-world scenario our EDA-Based PA can break protected implementations on several clone devices with the same power model.
\end{enumerate}

The paper is organized as follows, Sec.~\ref{sec.PA} briefly reviews the background on Profiling Attacks. The relevant work on automatic SCA, portable TAs and TAs on masked implementations are surveyed in Sec.~\ref{sec.Related}. The Estimation of Distribution Algorithms approach to automate the POIs selection in the SCA scenario is given in Sec. \ref{sec.EDAs}. Sec.~\ref{introAESPT} introduces the \textit{AES\_PT} dataset. Sec.~\ref{sec.EUCsingle} and Sec. \ref{sec.EUCport} contain the experimental results supporting our method. Finally, Sec.~\ref{sec.concl} concludes the paper. 

\section{Background on Profiling attacks}
\label{sec.PA}

Profiling attacks have become an archetype for SCA in recent years. The main idea for these attacks is to generate a model of the power consumption of a device, to be used for the recovery of sensitive information (i.e., cryptographic key). Therefore, these attacks consist of two phases: a profiling phase, in which the model is built out of a relatively big number of power traces, and an attack phase, in which the model is applied and the secret key is recovered with only a few traces. 
There exist different types of profiling attacks depending on the technique used for generating the model in the \textit{profiling phase}: the model can be generated by using standard classification techniques like in the first publications on Template attacks (TAs)~\cite{Chari2002template,Rechberger2005PracticalTA}; or Machine Learning (ML) techniques such as Support Vector Machine (SVM)~\cite{SVM1,SVM2,SVM3}, Random Forest (RF) \cite{RF} or recently introduced Deep learning (DL) ~\cite{Maghrebi2016BreakingCI,CNNagainstCM,CNNOV}; or even regression often called the Stochastic models approach~\cite{stochastic}. Nevertheless, ``classical'' TAs and ML are the two most popular approaches~\cite{TAvsML}, and in this paper, we focus on the former because it is a well-known and understood technique in the field of SCA. However, the approach remains valid in a broader context, not just for TA, although we pick TA to demonstrate results.

\subsection{Notation}
\label{subsec.Notation}

In this section we briefly describe the notation used throughout the document. In general we follow the notation proposed in \cite{BlueBook}, with some modification. The calligraphic letter $ \mathbf{T}$ denotes a set of traces $ \mathbf{t} $ (also called leakage vectors). In turn, each trace is formed by \textit{T} time samples $ \mathbf{t} = \left\lbrace t_{1},t_{2},\ldots,t_{T} \right\rbrace$. The total number of power traces $\mathbf{t}$ in a set of traces $ \mathbf{T}$ is denoted by $ |\mathbf{T}|$. We use $ v = f(p,k) $ for the targeted intermediate value, which is related to a public variable (usually the plaintext \textit{p}) and a cryptographic primitive (secret key \textit{k}). Calligraphic letter $ \mathbf{K}$ denotes the set of all possible keys, the secret key used by the criptographic algorithm (correct key) is denoted by $k^{*}$ and the total number of key hypotheses is denoted by $|\mathbf{K}|$. Regarding TAs, we denote each template by $h = (\mathbf{m},\mathbf{C})$. The probability that $x=i$ is denoted by $p(x=i)$. In the case of binary variables, we denote the probability that $x=1$ by $p(x=1)$ or simply $p(x)$. Finally, given a set of $ \mathbf{T}$ power traces, the attack outputs a key guessing vector $\textbf{g} = \left\lbrace g_{1},g_{2},...,g_{|\mathbf{K}|} \right\rbrace $ in decreasing order of probability. We understand Guessing Entropy ($ge$)~\cite{Standaert2009UnifiedFramework} as the average rank of $k^{*}$ in $\textbf{g}$ over multiple experiments. 

\subsection{Template attacks}
\label{subsec.TA}
Template attacks (TA) were proposed by Chari et al.~\cite{Chari2002template} for SCA as the first form of profiling attacks. They are based on building a multivariate model of the probability distribution of the leakage, which is commonly generated assuming that the leakages follow a Gaussian distribution (parametric estimation). In these attacks (and in SCA in general) it is common to work with power traces taken when the modeled device is handling some intermediate value $ v = f(p,k) $ related to a public variable (usually the plaintext \textit{p}) and the secret parameter (key \textit{k}). 

For that purpose, in the \textit{profiling phase} the attacker uses a set of $ n_p $ profiling traces ($ \mathbf{T}_{p}$) to build a Gaussian multivariate model, which is fully defined by a mean vector and a covariance matrix $(\mathbf{m},\mathbf{C})$~\cite{Chari2002template}, for each possible intermediate value $v$, creating the so-called \textit{templates}\footnote{Pairs of $(\mathbf{m},\mathbf{C})$, typically denoted by $h$}. Formally, the PDF which describes the multivariate normal distribution of a leakage vector $ \mathbf{t} = \left\lbrace t_{1},t_{2},\ldots,t_{T} \right\rbrace$ of length \textit{T} is described by Eq.(\ref{pdf}):

\begin{equation}\label{pdf} p(t \mid k)=\frac{1}{\sqrt{(2 \pi)^{T}\left|\mathbf{C}_{k}\right|}} \cdot \mathrm{e}^{-\frac{1}{2}\left(\mathbf{t}-\mathbf{m}_{k}\right)^{\prime} \mathbf{C}_{k}^{-1}\left(\mathbf{t}-\mathbf{m}_{k}\right)}
\end{equation}

Then, in the \textit{attack phase}, from a set of $ n_a $ attack traces ($ \mathbf{T}_{a}$) and its input data (plaintext), the attacker tries to guess the secret key. In order to do so, the attacker makes hypothesis about the secret key and calculate possible intermediate values $v_{i,j}$. Note that the intermediate values $v_{i,j}$ depend on the input data $ d_{i} (1 \leq i \leq D) $ and the key hypothesis $ k_{j} (1 \leq j \leq K) $, and hence each key hypothesis $ k_{j} $ suggest a template $(\mathbf{m},\mathbf{C})$ for each input value. 
Then, a \textit{discriminant score} $ D\left(k \mid \mathbf{t}_{i} \right) $ is computed for each key hypothesis $ k_{j} $ and the key hypothesis are ranked in decreasing order of probability.\footnote{In other words, the attack outputs a key guessing vector $\textbf{g} = \left\lbrace g_{1},g_{2},...,g_{|\mathbf{K}|} \right\rbrace $ in decreasing order of probability, as mentioned in Sec. \ref{subsec.Notation}.} Given a power trace $\mathbf{t}_{i}$, a commonly used discriminant derived from Bayes' is denoted by Eq.(\ref{disc}):

\begin{equation}\label{disc} D\left(k_{j} \mid \mathbf{t}_{i} \right)=p\left(\mathbf{t}_{i} \mid k_{j}\right) p(k_{j}) \end{equation}

This discriminant is obtained by omitting the denominator from Bayes' rule (see Eq. (\ref{bayes})), since is the same for each key hypothesis $ k_{j} $.

\begin{equation}\label{bayes}L\left(k_{j} \mid \mathbf{t}_{i}\right)=\frac{p\left(\mathbf{t}_{i} \mid k_{j}\right) p(k_{j})}{\sum_{k_{j}^{\prime}} p\left(\mathbf{t}_{i} \mid k_{j}^{\prime}\right) p\left(k_{j}^{\prime}\right)}\end{equation}

Moreover, if we assume that $p(k_{j})$ is an uniform probability, applying Bayes' rule is equivalent to computing the likelihood as in Eq.~(\ref{likelihood}):

\begin{equation}\label{likelihood}L\left(k_{j} \mid \mathbf{t}_{i}\right)=\frac{p\left(\mathbf{t}_{i} \mid k_{j}\right)}{\sum_{k_{j}^{\prime}} p\left(\mathbf{t}_{i} \mid k_{j}^{\prime}\right)}\end{equation}

\subsection{Point of Interest selection}
\label{POIselection}
Although TAs are optimal from an information-theoretic point of view, in their original formulation they pose a number of limitations, where computational complexity problems and the need for dimensional reduction the most critical ones~\cite{Choundary2018efficient}. The dimensionality reduction is commonly addressed in practice by selecting only a small number of the commonly big number of time samples of the power traces (Points of Interest (POI) selection \cite{Rechberger2005Practical}). In order to do so, the evaluator should select only the points in the power trace that contain the (most relevant) leakage, without losing other important information. This is typically done based on what we call ``POI selection graphics'' (See Fig. \ref{FSG_AESHD} and \ref{FSG_PLOT}). These graphics are obtained by applying certain functions (correlation, SOSD\cite{TAvsSTO}, SOST\cite{TAvsSTO}, SNR\cite{BlueBook}, ML-based \cite{Picek2019feature}) to power traces. Typically, a certain number of samples are selected in the higher values of those functions for building the templates.
There exist other techniques for reducing the number of samples in each power trace (often called compression methods), like Principal Component Analysis (PCA)\cite{Archambeau2006subspaces,Standaert2008Using} or Fisher's Linear Discriminant Analysis (LDA) \cite{LDA1,LDA2}, but here we focus on ``classical'' sample selection (for POI selection) as the most widely used technique in practice (without any lack of generalization). 

In addition, it should be noticed that the POI selection stage can be decisive and have a huge influence on the attack results, especially when we consider a portable scenario \cite{Rioja2020similarity}. Conversely, most of the related works start the analysis by assuming that the POIs are already pre-selected: Picek \textit{et al.} identified this problem and compared the performance of several feature selection techniques in a TA scenario~\cite{Picek2019feature}. However, our approach is different since the EDA-Based PA can be used together with this kind of techniques or just in a standalone way.

\section{Related Work}
\label{sec.Related}

\subsection{Automatic SCA}

In the field of SCA, there are only a few papers that discuss its automation. Moreover, as far as we know, no paper studies the automation of SCA in the context we are addressing in this paper: profiled attacks on cryptographic implementations. Specifically, the only papers that discuss automatic SCA are focused on using cache-timing attacks to automatically and more efficiently exploit complex Linux/Windows operating systems, which is a very different area of SCA.
For instance, in \cite{Gruss2015cacheTA} authors perform a novel automated attack in two stages on the T-table-based AES implementation of OpenSSL using cache-timing template attacks~\cite{Brumley2009cache}. Also, in \cite{Schwarz2019JavaScriptTA} Schwarz \textit{et al.} presented a fully automated approach to find subtle differences in browser engines caused by the environment and presented two new side-channel attacks on browser engines. They collected (automatically) all data available to the JavaScript engine to build templates. 
It should be noticed that, when we speak about ``templates'' or ``template attacks'' in the context of cache-timing attacks, we are referring to a different concept from the attacks covered in this paper. Cache template attacks were proposed in~\cite{Brumley2009cache}, and are named templates because they are inspired by Chari's work~\cite{Chari2002template}, as they have the same spirit in terms of performing the attacks in two steps (profiling and attack). In short, in the profiling phase, dependencies between the processing of secret information (e.g., specific key inputs or private keys of cryptographic primitives) and specific cache accesses are determined. Then, in the attack phase, secret values are derived based on observed cache accesses.

Nevertheless, our approach is different since we are addressing another concept of template attacks in a different field of SCA: we focus on an automatic search for the points in the power trace which will give us the best results with respect to where relevant leakage occurs. Thus, we claim that this approach can mitigate the need for a human in the loop of this kind of procedures since EDAs allow us to automate POIs tuning together with the template building and key recovering steps.

The biggest challenge for TA in particular, but also for PA in general, is the finding of proper time samples containing the leakage information (usually named Points Of Interest, POI). 
Thus, to ease the evaluation process in general, and TAs specifically, we propose to perform POIs tuning together with the template building and key recovering steps, automatically.
This allows expert evaluators to save time and parallelize tasks (improving the efficiency of the process) but also helps technicians without a deep knowledge of all the basics involved in these methods to implement TAs properly.

\subsection{Portability of Template Attacks}
\label{subsec.Portability}

The original idea of profiling attacks implies having two devices, the target device, and a clone hardware device which we can control completely for building the power consumption model. Practically, although this is not a realistic use case, traces for both profiling and attack phases are often captured from the same device in most of the related works on this topic~\cite{Chari2002template,Rechberger2005Practical,Archambeau2006subspaces,TAvsSTO,stochastic,TAvsML,kim2018noise,prouff2018ascad,Picek2019feature,SVM1,SVM2,SVM3,RF}. In fact, applying a power consumption model built with one device to a different copy (concept referred to as portability) is sometimes considered trivial, while it can be a challenging matter as some previous works have identified~\cite{Elaabid2012,FormalStudyPowerVariability,TAonDD,Bhasin2019MindTP,Rioja2020similarity}. The main reason is that there could still exist differences between ``identical'' devices or experimental setups. In practice, small variations in the construction of a device or aging can provoke behavioural deviations in the power consumption which can eventually lead to a failed attack. The same happens with subtle changes in the experimental setups used to acquire the traces: environmental changes, I/O interference, resonance due to LC and RC oscillators, influence of the past state, variations in the magnetic field penetration, etc.

To the best of our knowledge, there exist only a few papers on profiling attacks that consider portability~\cite{keyboardAES,Gohr2019CHES2S,cryptoeprint:2019:054}. In~\cite{Elaabid2012} the portability issue was identified and waveform realignment and acquisition campaigns normalization were proposed to improve the performance of portable TAs. In ~\cite{TAonDD,Choundary2018efficient} authors performed different attacks over four copies of the same device to argue on the impacts of the different copies. A more recent work~\cite{Bhasin2019MindTP} compares the performance of different machine learning techniques in the context of portable profiling attacks. In~\cite{Rioja2020similarity} authors propose an improved POI selection technique in order to address portability of data loading template attacks. They focused on finding points of common leakage in the power traces, avoiding the ones that perturb the model and make it limited to a particular copy of the device.

In this paper, we address the portability of TAs with EDAs in two use cases (Sec. \ref{UC3} and \ref{UC4}) and using the new aforementioned \textit{AES\_PT} dataset, on which we are able to break a masked AES implementation in four clone devices with the same probabilistic model.

\subsection{Template Attacks on Masking}

In \cite{Oswald2007template}, Oswald \textit{et al.} discussed different types of TAs on masked AES software implementations on an 8-bit microcontroller. They applied two types of attacks: TAs combined with second-order techniques and template-based DPA attacks with extra calculation considering the masks, see Eq.~(\ref{maskingP}):

\begin{equation}\label{maskingP}p\left(\mathbf{t}_{i} \mid k_{j}\right)=\sum_{m=0}^{M-1} p\left(\mathbf{t}_{i} \mid k_{j} \wedge m\right) \cdot p(m)
\end{equation}

Where $\mathbf{t}_{i}$ represents a power trace, $\mathbf{k}_{j}$ represents a key hypothesis, and $\mathbf{m}$ represents the different values the mask can take (being $M$ the maximum number of different values that the mask can take). 

Regarding the attacks, all of them break the implementation (with a different number of traces), concluding that, in the scenario of TAs, there is no difference in the security of masked and unmasked implementations. The best attack is claimed to be the template-based DPA, which can recover the key from about 15 traces (using 10k traces in the \textit{profiling phase}).

Later, several works have performed ``regular'' template-based DPA attacks (i.e. without the extra calculation considering the masks), with successful results ~\cite{kim2018noise,prouff2018ascad}. On the other hand, since in masking implementations POI selection graphics are not conclusive because the intermediate value is randomized (masked), the POI selection step is usually performed by using DPA attacks~\cite{Oswald2007template} or using PCA/LDA~\cite{kim2018noise,prouff2018ascad}. In this work, we follow a different approach since the POI selection is done intrinsically in our EDA-Based PA, as explained below.

\section{The search of points of interest by means of randomized optimization heuristics}
\label{sec.EDAs}

As mentioned above, the goal of using randomized optimization heuristics is to automate POIs tuning together with the template building and key recovering steps. The idea is to search in the space shaped for all possible groupings of POIs in the power trace for the best ones, that is, the POIs that turn attacks into reality. However, an exhaustive enumeration of every combination of points of the power traces is exponential, that is, if the number of points is $T$, the number of possible
combinations is $2^T$ which makes an exhaustive enumeration impractical with only dozens of points. In order to accomplish this kind of tasks, optimization heuristics have gained popularity due to their high efficiency in finding optimal solutions in complex problems where other exact methods take too much time and CPU resources.

In the field of optimization heuristics, several approaches can be used including genetic algorithms among others. In this work, a new search strategy is proposed that is based on a quality measure combined with recent efficient evolutionary computation algorithms namely, estimation of distribution algorithms.

\subsection{Estimation of distribution algorithms}
Estimation of distribution algorithms (EDAs) are a novel class of evolutionary optimization methodology that was developed in the last decade as a natural alternative to genetic lgorithms~\cite{Muhlenbein1996edas,Larranaga2002edas,Pelikan2003hierarchical,Lozano2006edas,Coffin2008eda}.

EDAs show different advantages over genetic algorithms such as the absence of multiple parameters to be tuned (e.g. crossover and mutation probabilities) or the expressiveness and transparency of the probabilistic model that guides the search. This novel class of evolutionary optimization methodology has been proven to be better suited than GAs in some applications achieving competitive and robust results in the majority of tackled problems.

\subsubsection{Introduction\\}
\begin{figure}[!h]
	\centering
	\includegraphics[width=0.9\textwidth]{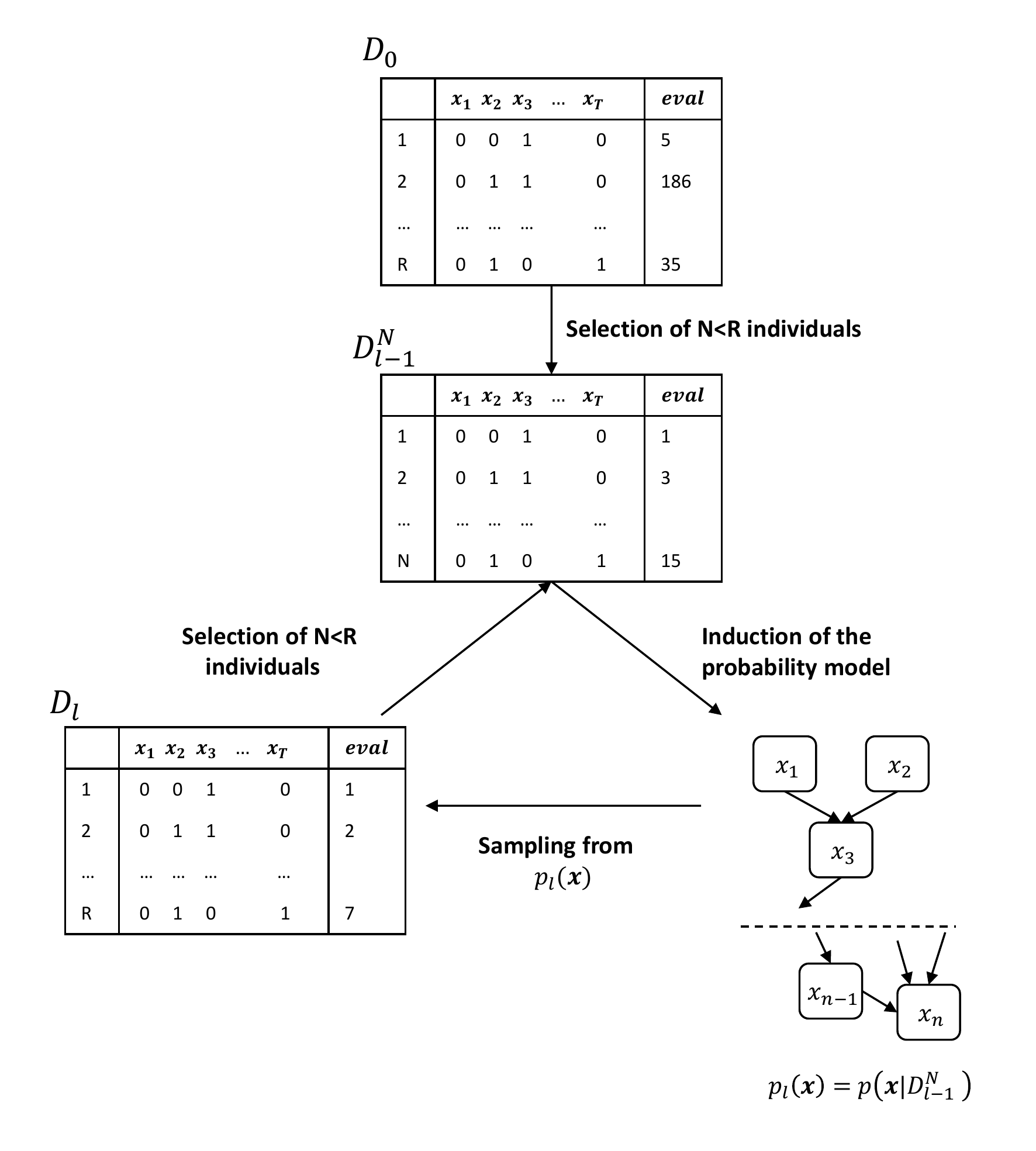}
	\caption{\label{fig:iIlustrationEDA} Illustration of EDA process}
\end{figure}

EDAs as the GAs work with a population of candidate solutions (in our case candidate points of oints). Fig.~\ref{fig:iIlustrationEDA} shows general schematics for any EDA approach. 

Initially, a random sample of candidate groups of POIs is generated. These candidate POIs are evaluated by means of an objective function. According to this evaluation, the best points are selected. Then, the selected solutions are used to learn a probabilistic model, and based on this new model a new set of groups of POIs is sampled. The process is iterated until the optimal value has been found or another termination criterion is fulfilled.

\subsubsection{A basic taxonomy of EDAs\\}
The literature shows a variety of models and learning algorithms, the selection of the best model for a given problem is not always straightforward. A criterion is to balance the computational cost of learning with respect to the complexity of the probabilistic model. Both aspects are strongly related to the problem dimensionality (i.e. the number of variables) and to the type of variable (e.g. continuous, discrete, mixed). 

Researchers should note that simple models have low requirements in different aspects such as computational resources or the complexity of the learning process. Whereas the complex models are able to represent more complex relationships but requiring more sophisticated data structures and costly learning processes. 
Depending on the problem researchers should take into account the balance between the search efficiency and the cost associated with the chosen strategy. An additional criterion that should be taken into account is the previous knowledge and to chose a probabilistic model to represent this knowledge.

In order to help the researcher to find a suitable EDA type, these can be classified according to the dependencies between the variables as follows:
\begin{itemize}
    \item   \textbf{Univariate EDAs}, assume that all variables are independent and therefore the joint probability can be factorized as a product of univariate marginal probabilities. An example of this kind of models is the Univariate Marginal Distribution Algorithm (UMDA)~\cite{Muhlenbein1997equation,Armananzas2008review,Gonzalez2002mathematical}. These algorithms are the simplest EDAs with the best CPU performance in terms of time.
    \item   \textbf{Bivariate EDAs} represent low complexity interdependencies between variables, examples of this class are mutual information maximization for input clustering (MIMIC)~\cite{DeBonet1996many}, the bivariate marginal distribution algorithm BMDA~\cite{Pelikan1998bivariate}, dependency tree-based EDAs ~\cite{Baluja1997UsingOD} or the tree-based estimation of distribution algorithm(Tree-EDA)~\cite{Santana1999edge}.
    \item   \textbf{Multivariate EDAs} factorize the joint probability distribution considering statistics of order greater than two. The complexity of the model as well as the effort required to estimate the parameters that best suit the selected points is significantly greater. Algorithms that belong to this group are EBNA~\cite{Baluja1997UsingOD}, and BOA~\cite{Pelikan99boa}.
\end{itemize}
For additional and detailed information regarding the different models that constitute the family of EDAs, see~\cite{Larranaga2002edas,Pelikan2002survey}.

\subsection{EDAs in a SCA scenario}
\label{subsec.EDA-SCA}
In order to explain how EDAs can be applied in the SCA scenario, we will focus on an AES implementation, same as in the experimental use cases described in Sec.~\ref{sec.EUCsingle} and Sec.~\ref{sec.EUCport}. 
Given a set of $ \mathbf{T} $ power traces (with $T$ samples per trace) labeled with an 8-bit value corresponding with the processed intermediate value (as typical for SCA on AES implementations), the problem consists of selecting the best points of interest (samples) for building templates in a TA scenario~\cite{Chari2002template,Rechberger2005Practical}. 

This task can be performed efficiently by using an EDA which explicitly represents each of the elements involved in the problem, in this case the power traces. Our approach is to consider a vector of binary variables (candidate points of interest) of length $T$:

$$ \textbf{x} = \lbrace x_{1}, x_{2}, \ldots, x_{n}, \ldots, x_{T} \rbrace = (0, 1, 0, \ldots, 0) $$

\begin{figure}[!h]
	\centering
	\includegraphics[width=1\textwidth]{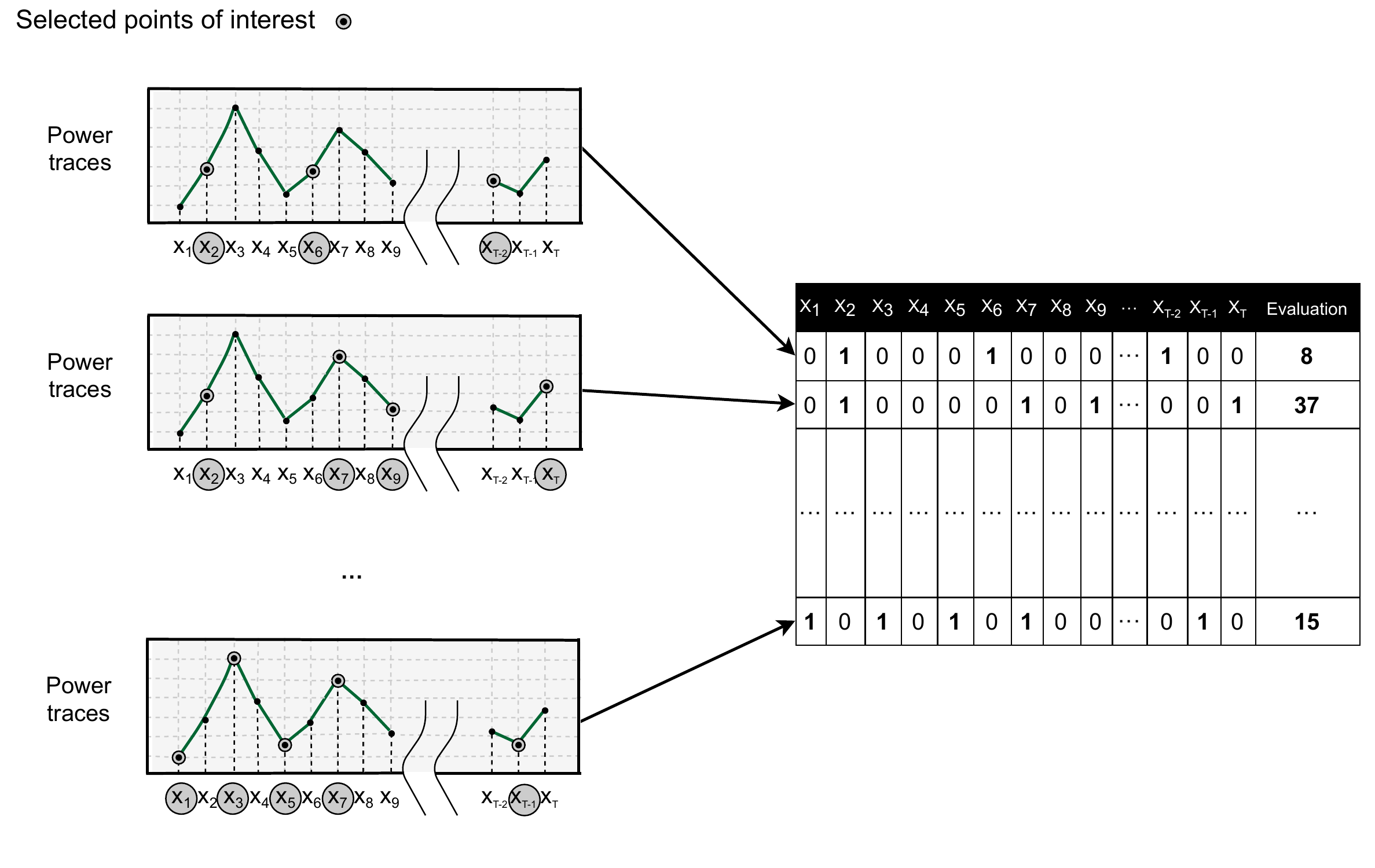}
	\caption{\label{fig:EDASCAScenario} EDA-SCA Model.}
\end{figure}

The cardinality of the search space for this problem is $2^{T}$. Each binary variable  $ \lbrace x_{1}, x_{2}, \ldots, x_{T} \rbrace $ corresponds to one time sample of the power traces $ \left\lbrace t_{1},t_{2},\ldots,t_{T} \right\rbrace$. If the value of a certain binary variable is 1 ($ x_{n} = 1 $), that means that the corresponding sample of the power traces $ t_{n} $ will be selected for building the templates, and vice versa. This process is depicted in Figure \ref{fig:EDASCAScenario}.

The selection of points of interest requires the handling of hundreds or thousands of variables which involves handling complex models and therefore the CPU time and memory required can be huge. Furthermore, the exponential nature of the problem represents a limitation when the number of traces increases when necessary.

The decision of selecting complex models where relationships between power traces are explicitly represented can have relevant disadvantages such as:  

\begin{itemize}
    \item   The first one is associated with the complexity of the model when representing points of interest. If each point of the initial population must represent a codified set of points of interest then it involves the use of a more complex probabilistic model. As a consequence the different parts are affected as follows:
            \begin{itemize}
                \item   The size of the population. In order to correctly estimate the parameters of the model, there should be a minimum number of instances that increases with the complexity of the model. This represents a significant increase in time and computational resources.
                \item   The sampling of the model. An increase in complexity necessarily involves an increase in time and CPU resources causing the sampling of the model that can be prohibitive.
                \item   The learning of the probabilistic model. The existence of a more complex model causes that more computations are required to correctly estimate the parameters.
                \item   Computation of the measure. With bigger populations, the increase of the computation time is very significant but can also be a limiting factor. 
            \end{itemize}
    \item   Finally, our proposed algorithm performs a search in a huge space that is exponential in the number of variables to be considered. Any kind of increase in the number of variables has a great impact on the difficulty of finding good promising POIs candidates.
\end{itemize}

Therefore, due to the prohibitive costs cited previously the use of complex models between points of interest is discarded. Based on this premise, EDAs such as UMDA are suitable candidates to handle the problem of selection of points of interest. The model chosen is known as univariate marginal distribution algorithm for discrete domains: $\textit{UMDA}_{d}$. This method considers a model where 
there are no interrelations between the variables, and the probability distribution can be learnt as:

\begin{equation}
    p_{l}(\boldsymbol{x})=\prod_{i=1}^{T}p_{l}(x_{i})
\end{equation}

This distribution that represents the product of $T$ independent probability distributions associated with the points of interest $p_{l}(x_{i})$ is a Bernoulli probability distribution that takes on two values: 
$1$ if the point is selected and $0$ if the point is not selected in the considered candidate vector. Based on this model, the estimation of the parameters of the model is performed based on marginal frequencies of the selected subset of points of interests:

\begin{equation}
    p_{l}(x_{i})=\frac{\sum_{j=1}^{T}\delta_{j}(X_{i}=x_{i}|D_{l-1}^{Se})}{T}
\end{equation}

where:

\begin{equation}
    \delta_{j}(X_{i}=x_{i}|D_{l-1}^{Se})=\begin{cases} \mbox{1,} & \mbox{if in the $j$-th case of} D_{l-1}^{Se},X_{i}=x_{i} \\ \mbox{0,} & \mbox{otherwise} \end{cases}
\end{equation}

\subsection{Description of the algorithm}
Thus, since EDAs are based on building explicit probabilistic models of promising candidates, each binary variable $x_{n}$ has its own probability of being one $p(x_{n}=1) = p(x_{n})$, which will be (re)computed after each iteration. Formally speaking, after each iteration, we estimate the probability distribution $p(\mathbf{x})$ of promising candidates from the highest quality solutions. 
The proposed algorithm works as follows:

\begin{enumerate}
    \item Firstly, the initial population $D_{0}$ of $R$ individuals is generated. This is usually performed by assuming a uniform distribution on each variable (but can also be generated manually). In other words, based on the probabilities of each binary variable of being one $p(x_{n})$, we sample an initial population of $R$ individuals (POI selection candidates). Whenever possible, we propose to sample the initial population based on a POI selection graphic, as detailed below (Sec. \ref{EDA&FSS}). Then, each individual is evaluated: a TA is performed with the POI selection candidate, quantifying the success of the attack with a proper evaluation function. This evaluation function has to be set by the evaluator and will depend on the desired result. We propose two different evaluation functions, based on the Guessing Entropy of the attacks, depending on the use case (Eq. (\ref{eq.uc1}) and (\ref{eq.uc3})), as explained in Sec. \ref{UC1.ad}, Sec. \ref{UC2.ad}, Sec. \ref{UC3.ad} and Sec. \ref{UC4.ad}.
    
    \item Secondly, the initial population is ranked using the evaluation results in order to select a number $N$ ($N<R$) of individuals from the previous $D_{l-1}$ population and evolve towards the next one $D_{l}$. We understand $D_{l-1}^{N}$ as the set of $N$ selected individuals from the generation $l - 1$.
    
    \item Thirdly, the $T$-dimensional probabilistic model which represents interdependencies between the $T$ variables is inducted. This is also known as the learning procedure and is the most critical part of the process. As usual in practice, we start considering the simplest case: variables are independent (no conditional probabilities). In other words, we (re)compute each probability $p(x_{n})$ based only on the number of times that $ x_{n} = 1 $ in $D_{l-1}^{N}$. Formally speaking,  $p_{l}(x_{n}) = p(x_{n}|D_{l-1}^{N})$.
    
    \item Finally, the new population ($D_{l}$) of $R$ new individuals is generated by simulating the probability distribution obtained in the learning process (Step 3). As in the second step, the individuals are evaluated and the best set of individuals is selected. Steps 2, 3, and 4 are repeated until a stopping condition is achieved (reaching a certain number of iterations, uniformity in the generated population, etc.). 
    
\end{enumerate}

\subsection{Improving EDA-Based PA with POI selection techniques}
\label{EDA&FSS}
The initial population of the EDA-Based PA can be generated from a uniform distribution (e.g., Bernoulli distribution) or can be generated in a more specific manner. We propose to sample the initial population from POI selection graphics (Sec. \ref{POIselection}) in order to improve the performance of the EDA. It should be noticed that this approach can not be always followed. For instance, in a masked implementation, POI selection graphics are not conclusive since the intermediate value is randomized (masked). Nevertheless, where applicable, the probability of each binary variable $x_{n}$ of being one ($p(x_{n}=1) = p(x_{n})$) can be obtained from Eq.(\ref{eq1}).

\begin{equation}\label{eq1}
    p^*(x_{n}) = \alpha \cdot p(x_{n})
\end{equation}

Where $p(x_{n})$ is a default probability value (i.e. Bernoulli distribution with $p(x_{n}) = 0.5 $) and $\alpha$ is computed by:

$$ \alpha(n) = \frac{s_{n} - s_{Min}}{s_{Max} - s_{Min}}; \quad \alpha \, \epsilon \, [0,1]$$

Here, $s_{n}$ is the corresponding sample of the POI selection graphic, and $s_{Max}$ and $s_{Min}$ the maximum and minimum value of the graphic respectively. By doing this, we will accelerate the search since we are considering the points with higher value in the POI selection graphic in most of the individuals (they have a greater probability) and avoiding the samples which do not represent leakage (points with small correlation, SNR, SOST, etc. will have the lowest probabilities).

\section{Experimental use cases: Single Device}
\label{sec.EUCsingle}

In this section, we demonstrate the suitability of our method in a TA scenario (using the same device for both the profiling and attack phase) with two use cases, an unprotected AES implementation and a protected AES implementation. 

It should be noticed that, as mentioned above, the proposed technique is suited for the context of side-channel evaluation. In this context, the strongest possible capabilities are commonly assumed for the attacker. That implies, among other aspects, that the attacker has physical access to the DUT and the capability to send a large number of chosen messages to the device. We also assume that the attacker knows the input and output data. In addition, it is likely that the attacker has some limited information about the internal workings of the device. These assumptions are made for all the experimental use cases proposed in this work.

\subsection{Unprotected AES implementation on FPGA (AES\_HD)}
\label{UC1}

In this use case, we showcase the applicability of EDAs in a simple TA scenario (unprotected cryptographic implementation, not considering portability). Although straightforward, the present use case represents a clear example of how EDAs can be applied in the SCA field. Nevertheless, more complex experiments are considered in the remaining use cases. 

\subsubsection{Target description: AES\_HD\\}
The \textit{AES\_HD} dataset includes 100\,000 traces from an unprotected implementation of AES-128 on FPGA. The AES core is written in VHDL in a round-based architecture, taking 11 clock cycles for each encryption. The core is wrapped around by a UART module in order to enable external communication. The module is designed to allow accelerated measurements to avoid any DC shift caused by environmental perturbations. The design was implemented on Xilinx Virtex-5 FPGA of a SASEBO GII evaluation board with a total area of 1850 LUT and 742 flip-flops. Side-channel traces were obtained by measuring the electromagnetic radiation produced by a decoupling capacitor (on the power line) with a high sensitivity near-field EM probe and a Teledyne LeCroy Waverunner 610zi oscilloscope. A suitable and commonly used leakage model when attacking the last round of an unprotected hardware implementation is the register writing in the last round \cite{DPAcontest}:

\begin{equation}
Y\left(k^{*}\right)=H W(\underbrace{\mathrm{Sbox}^{-1}\left[C_{b_{1}} \oplus k^{*}\right]}_{\text {previous register value }} \oplus \underbrace{C_{b_{2}}}_{\text {ciphertex t byte}})
\end{equation}

\subsubsection{Attack details\\}
\label{UC1.ad}

Table \ref{uc1.details} gives the details of the attack. It should be noticed that, as usual, for the profiling phase we are using 10\,000 random traces of the device processing different intermediate values, i.e. all the 256 possible values $v$ can take since it is an 8-bit value. On the other hand, for the attack phase, we are using 300 traces of the device processing the same $v$. These attack traces are obtained from the dataset by selecting at random 300 traces of the device processing a certain $v$ value, chosen at random too. Nonetheless, this is done on purpose since although this implies that our use case is ``simpler'' than other related works ~\cite{kim2018noise,Picek2019feature} (in which they perform the attack phase with 25\,000 traces with different associated $v$ values), it represents a clear and straightforward example that serves to explain our approach. Moreover, this use case indeed corresponds to a ``realistic'' TA scenario. In theory, in a ``realistic'' TA we can obtain profiling traces from all possible $v$ values because we are using a clone device that we can fully control. However, in such a scenario we cannot fully control the attacked device, which has its own secret key that we want to recover. Therefore, we can only obtain a small number of attack power traces in which the cryptographic key is fixed because it is linked to the device. In this particular case, the plaintext also would be fixed, thus the intermediate value is always the same, which could correspond to a use case in which the device sends the same encrypted value several times. In any case, more demanding use cases have been considered in the rest of the experiments. 

\begin{table}[!h]
    \centering
    \begin{tabular}{cc}
        \hline
        Variable & Value and description \\
        \hline
        TA Details & Variance only, no pooling \\
        Power Model & IV model \\
        \# profiling traces & 10\,000 \\
        \# attack traces & 300 \\
        Correction Factor & 10 \\
        Population size & 20 \\
        \# Iterations & 10 \\
         \hline
    \end{tabular}
    \caption{Attack specifications (Unprotected AES)}
    \label{uc1.details}
\end{table}

Regarding EDAs, even though they have less amount of parameters involved than other searching techniques, we still have to adjust the number of iterations and the population size. Also, we have to select a proper score function which enables us to correctly evaluate each individual of the entire population. In this case, we consider the following formula to evaluate each individual (POI candidate):

\begin{equation}
\label{eq.uc1}
Eval(ind) = \left \{ \begin{matrix} - CF*\frac{ge}{256} & \mbox{if }ge\neq\mbox{1}\\ - CF*\frac{n_{POI}}{n_{samples}}*\frac{ge}{256} & \mbox{if }ge=\mbox{ 1}\end{matrix}\right.  
\end{equation}

\begin{figure}[!h]
    \centering
    \includegraphics[width=0.7\textwidth]{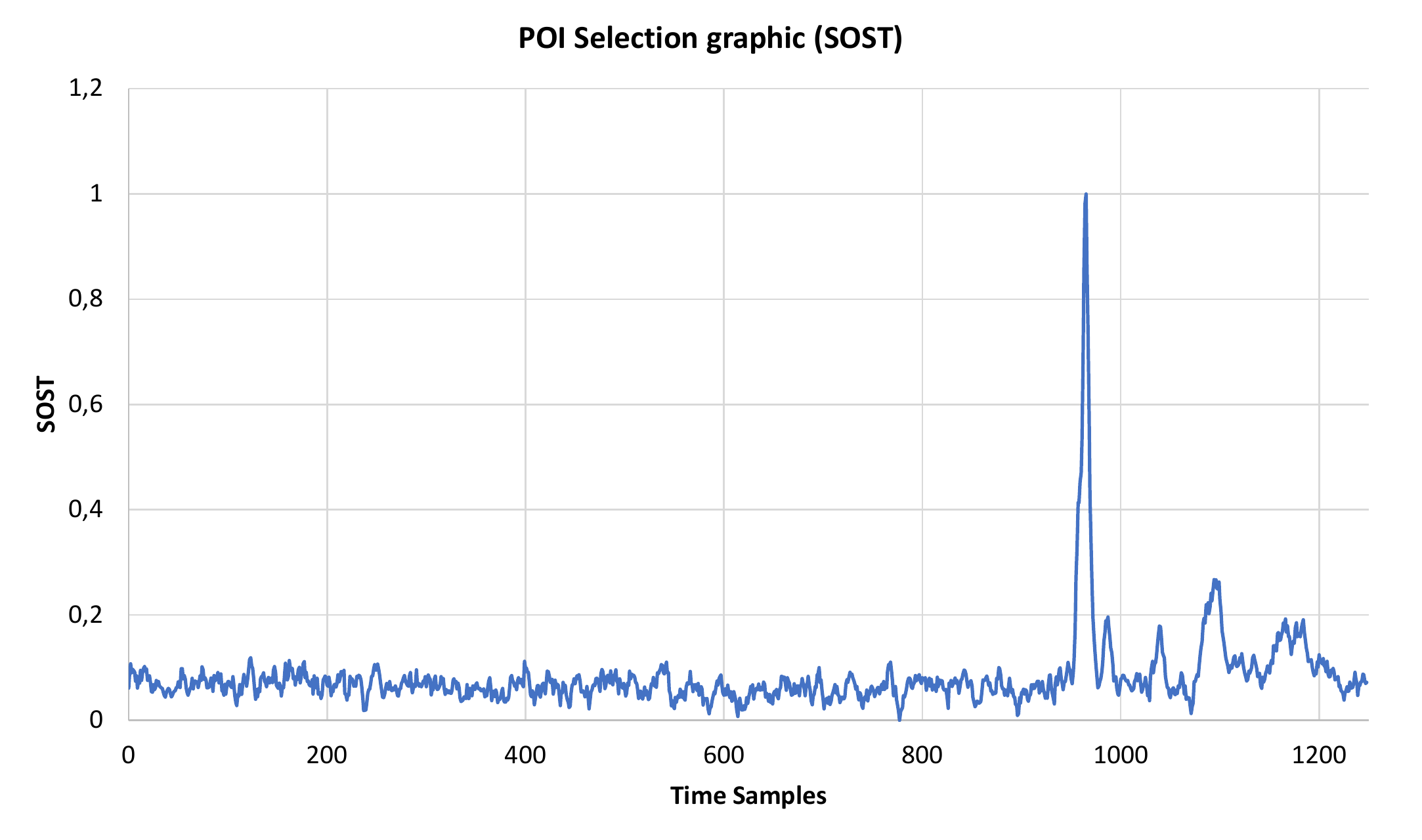}
    \caption{POI selection graphic (SOST) of the targeted intermediate value, normalized between 0 and 1}
    \label{FSG_AESHD}
\end{figure}

Where $ge$ is the result of attacking the attack (guessing entropy; rank of the correct candidate). The negative sign is there in terms of minimizing the ranks of the correct candidates, $n_{POI}$ is the number of POI used to build templates (to guide the EDA into minimizing the number of POI if the attack is successful) and $n_{samples}$ is the number of samples per trace. We also use a correction factor ($CF$) to accelerate the search. We divide $ge$ by 256 (maximum guessing entropy in this case) in order to bound the value between 0 and 1, and the same happens with $n_{POI}$ and $n_{samples}$.
In addition, we are using a POI selection graphic (Fig. \ref{FSG_AESHD}) obtained from the profiling traces to guide the search even more (as explained in Sec. \ref{EDA&FSS}).

\begin{figure}[!h]
\begin{minipage}{0.47\textwidth}
\centering
\begin{tabular}{cccc}
\hline
 & Factor name & Low setting & High setting\\
\hline
A &  CF &  1 & 10 \\
B &  \# Iter. & 5 & 10 \\
C &  Pop. Size & 10 & 20 \\
\hline
\end{tabular}
\captionof{table}{Working variables and values (DoE)}
\label{ParameterSelectionAESHD}
\end{minipage}\hfill
\begin{minipage}{0.47\textwidth}
\centering
\includegraphics[width=1\textwidth]{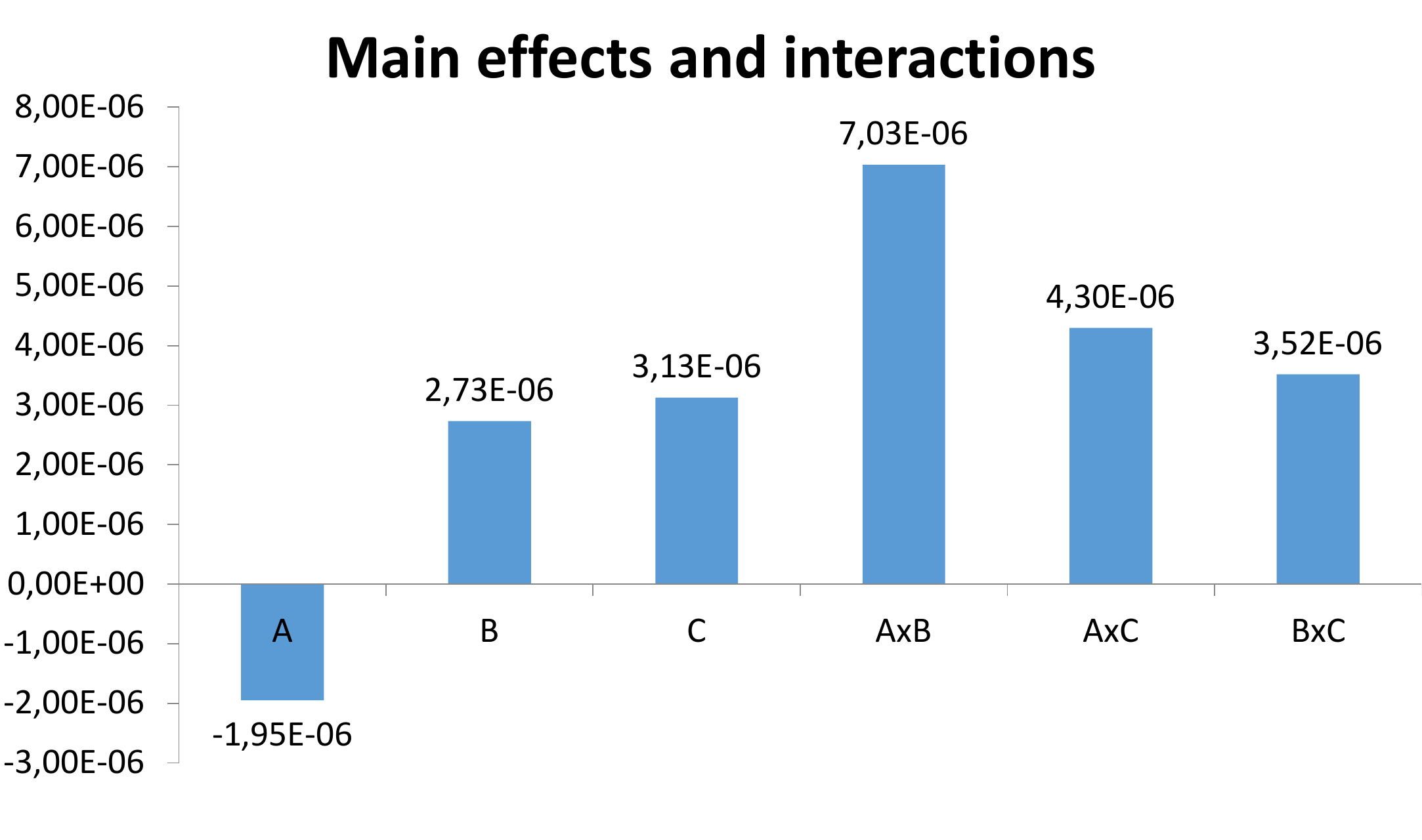}
\captionof{figure}{\label{DOE_AESHD} Effects and Iterations (DoE)}
\end{minipage}\hfill
\vspace{0.5cm}
\begin{minipage}{1\textwidth}
\centering
\begin{tabular}{ccccccc}
\hline
Exp & A & B & C & $ n_{POI} $ & $ge$ & \textbf{Eval}\\
\hline
1 & 1 & 5 & 10 & 31 & 1 & -4.84E-5 \\
2 & 1 & 5 & 20 & 41 & 1 & -6.41E-5 \\
3 & 1 & 10 & 10 & 45 & 1 & -7.03E-5\\
4 & 1 & 10 & 20 & 38 & 1 & -5.94E-5 \\

5 & 10 & 5 & 10 & 52 & 1 & -8.13E-5\\
6 & 10 & 5 & 20 & 43 & 1 & -6.72E-5\\
7 & 10 & 10 & 10 & 40 & 1 & -6.25E-5\\
\textbf{8} & \textbf{10} & \textbf{10} & \textbf{20} & \textbf{30} & \textbf{1} & \textbf{-4.69E-5}\\
\hline
\end{tabular}
\captionof{table}{Table of results (DoE)}
\label{ExperimentRoundsAESHD}
\end{minipage}\hfill
\end{figure}

With this setup, we have essentially three variables (factors) to adjust: the correction factor, the number of iterations, and the population size (Table \ref{ParameterSelectionAESHD}). Although it is not mandatory, we recommend using Experimental Design \cite{Tanco2009doe} (also known as Design of Experiments or DoE) in order to find proper values for the EDA's parameters, same as in \cite{ePrintSixSigma,Paguada2020controlling}. The idea is to identify and quantify the effect each variable has in the experimental results, which is especially useful when the evaluator is not sure about how to select appropriate values. Results of our experimental design are represented in Fig. \ref{DOE_AESHD} and Tables \ref{ParameterSelectionAESHD} and \ref{ExperimentRoundsAESHD}.

In a nutshell, we perform $2^3=8$ experiments with all possible combinations of variables A,B and C and their low and high settings (Tables \ref{ParameterSelectionAESHD} and \ref{ExperimentRoundsAESHD}). Then we apply the DoE formula and compute the effect of each variable and their interactions in the output of the attack (Fig. \ref{DOE_AESHD}). 



It should be noticed that the attack is successful in all the experiments (Table \ref{ExperimentRoundsAESHD}, column $ge$, the rank of the correct candidate is one for all the column). Thus, although selecting proper values for the three variables in this use case is quite straightforward, it is a good ``toy example'' for showcasing how to apply the design of experiments and interpret the effect each variable has in the results. For instance, if we observe Fig. \ref{DOE_AESHD}, we can see that variables B and C have a positive effect (the experimental results are better with the high setting). On the other hand, variable A has a negative effect (the experimental results are better with the high setting). Note that this is true in general for all the experiments but the best result is experiment 8 (all variables in its high setting) and hence we select that combination of variables for this use case. For a more detailed explanation of the procedure, we refer to \cite{ePrintSixSigma,Paguada2020controlling}.


\subsubsection{Results\\}

In Table \ref{uc1.ITER1}, results from the first iteration are shown. In the left part of the table we can observe two columns: Individual and Evaluation result while in the right part of the table we can see detailed information of the TA: number of POI used and performance of the portable TA (Guessing Entropy; Average rank of the correct key candidate). Please note that we have not represented some of the individuals in order to reduce the tables' size.

\begin{figure}[!h]
    \centering
    \includegraphics[width=1\textwidth]{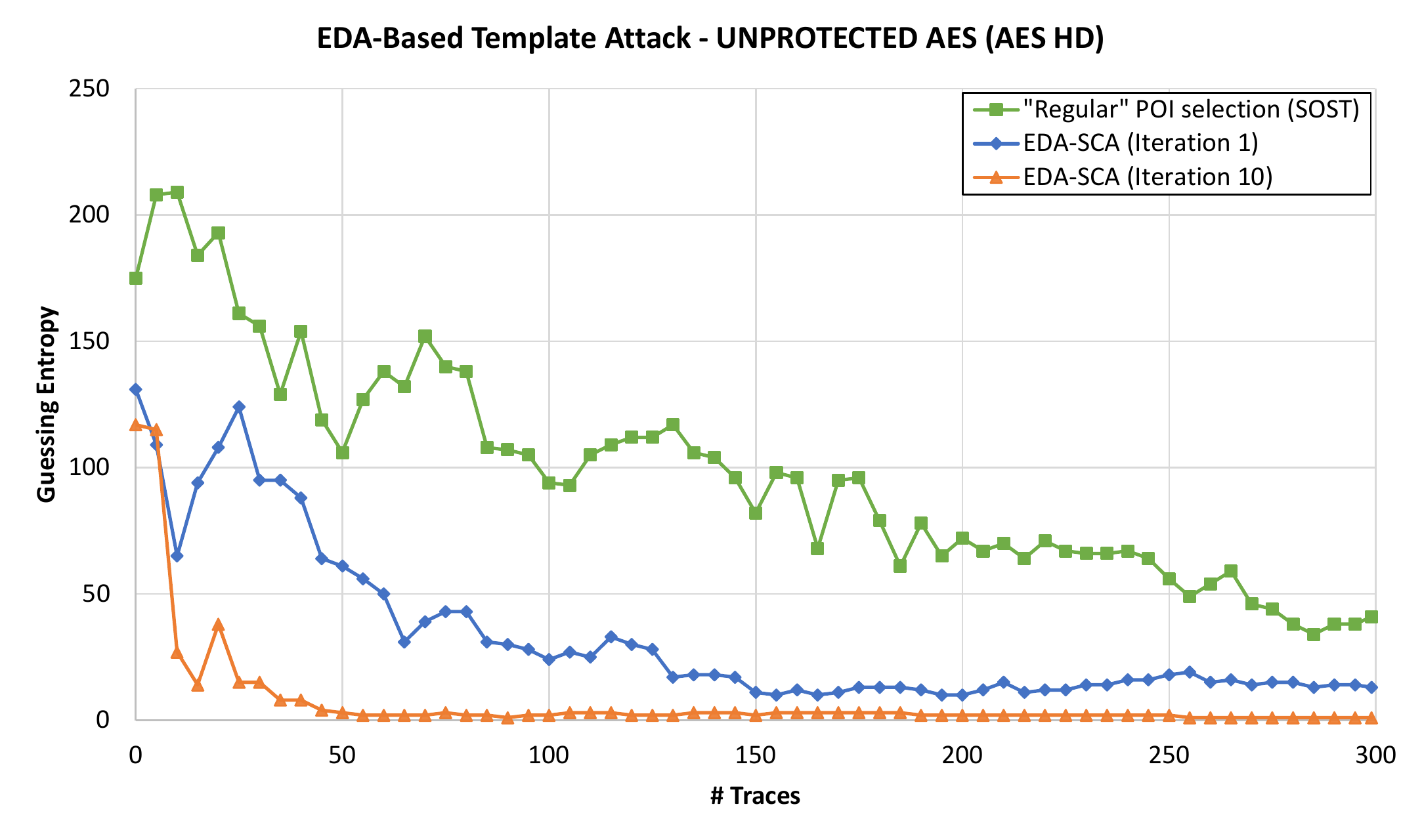}
    \caption{Unprotected AES Implementation (Guessing Entropy)} 
    \label{fig:geAESHD}
\end{figure}

The process is repeated until we reach the last iteration (iteration 10), represented in Table \ref{uc1.ITER10}. If we compare the results of the first and last iteration we can observe that whereas in the first iteration the results of the portable attacks are quite poor, in the last round we succeed in the attack with almost all the individuals. 
A more graphical representation of the obtained results can be observed in Fig. \ref{fig:geAESHD}. There you can see the experimental results (guessing entropy) of using a ``regular'' POI selection (just select 20 time samples on the highest SOST values), the results of the best individual (POI selection candidate) of the first iteration, and the results of the best individual (POI selection candidate) of the last iteration. It should be noticed that the ``regular'' POI selection throws the worse results (rank of the correct candidate is 41 after 300 traces). On the other hand, both EDA-Based PA POI selections have better performance: We obtain a rank of 13 and 1 after 300 traces respectively. In conclusion, this experimental use case serves to show how our method optimizes the attacks until an optimal result is achieved (the rank of the correct candidate is one after the attack), improving the results obtained with a ``regular'' POI selection with almost no effort.

\begin{table}[!h]
\begin{minipage}{0.47\textwidth}
\centering
	\begin{tabular}{cccc}
	\hline
	Ind  & Eval& $ n_{POI} $ & $ge$ \\
	\hline
    1 & -5.08E-01 & 43 & 13 \\
    2 & -9.77E-01 & 54 & 25 \\
    3 & -1.054688 & 48 & 27 \\
    4 & -1.054688 & 45 & 27 \\
    5 & -1.367188 & 47 & 35 \\
    6 & -1.367188 & 40 & 35 \\
    7 & -1.406250 & 44 & 36 \\
    8 & -1.445313 & 43 & 37 \\
    9 & -1.757813 & 58 & 45 \\
    10 & -1.914063 & 50 & 49 \\
    11 & -2.265625 & 38 & 58 \\
    12 & -2.578125 & 53 & 66 \\
    13 & -2.578125 & 40 & 66 \\
    14 & -2.578125 & 40 & 66 \\
    15 & -2.968750 & 56 & 76 \\
    16 & -3.320313 & 47 & 85 \\
    17 & -3.945313 & 50 & 101 \\
    18 & -4.062500 & 53 & 104 \\
    19 & -4.062500 & 56 & 104 \\
    20 & -4.335938 & 40 & 111 \\
	\hline
	\end{tabular}
	\caption{Results of the first iteration}
	\label{uc1.ITER1}
\end{minipage}\hfill
\begin{minipage}{0.47\textwidth}
    \centering
	\begin{tabular}{cccc}
	\hline
	Ind & Eval & $ n_{POI} $ & $ge$ \\
	\hline
    1 & -4.69E-04 & 30 & 1 \\
    2 & -5.31E-04 & 34 & 1 \\
    3 & -5.47E-04 & 35 & 1 \\
    4 & -5.47E-04 & 35 & 1 \\
    5 & -5.63E-04 & 36 & 1 \\
    6 & -5.63E-04 & 36 & 1 \\
    7 & -5.94E-04 & 38 & 1 \\
    8 & -5.94E-04 & 38 & 1 \\
    9 & -6.09E-04 & 39 & 1 \\
    10 & -6.09E-04 & 39 & 1 \\
    11 & -6.25E-04 & 40 & 1 \\
    12 & -6.41E-04 & 41 & 1 \\
    13 & -6.41E-04 & 41 & 1 \\
    14 & -6.56E-04 & 42 & 1 \\
    15 & -6.72E-04 & 43 & 1 \\
    16 & -7.81E-02 & 35 & 2 \\
    17 & -7.81E-02 & 36 & 2 \\
    18 & -7.81E-02 & 39 & 2 \\
    19 & -7.81E-02 & 40 & 2 \\
    20 & -7.81E-02 & 37 & 2 \\
	\hline
	\end{tabular}
	\caption{Results of the last iteration}
	\label{uc1.ITER10}
\end{minipage}\hfill  
\end{table}

\subsection{Protected AES implementation}
\label{UC2}
In this use case, we showcase the applicability of EDAs in a more challenging scenario: a protected AES implementation. In order to demonstrate that our approach can improve on state-of-the-art results, we are again using a freely available dataset ASCAD~\cite{prouff2018ascad}.

\subsubsection{Target: ASCAD Dataset\\}
ASCAD was presented in \cite{prouff2018ascad} as the first open database for DL-SCA. The target platform in this data set is an 8-bit AVR microcontroller (ATmega8515), implementing a masked AES-128 cipher \cite{BlueBook} and the traces are obtained by measuring the electromagnetic emanation of the device. The data set provides 60\,000 traces where 50\,000 are used for profiling and 10\,000 for the attack. These traces are a window of 700 relevant raw samples per trace, representing the third byte of the first round masked Sbox operation. For a deeper explanation of the ASCAD dataset, we refer to \cite{prouff2018ascad}. As the sensitive intermediate value we use an Sbox output:
\begin{equation}
Y^{(i)}\left(k^{*}\right)=\operatorname{Sbox}[P_{3}^{(i)} \oplus k^{*}]  
\end{equation}

\subsubsection{Attack details\\}
\label{UC2.ad}

Table \ref{uc2.details} summarizes the specifications of the attack. We are collecting 50\,000 random traces for building the templates (profiling subset) and 300 random traces for the attack (attack subset). For the evaluation function, we are using Eq.~(\ref{eq.uc1}), as in the previous use case. In this case, since it is a masked implementation and obtaining the POI graphs is not straightforward, we sample the initial population from a Bernoulli distribution with $ p = 0.1 $. 

\begin{table}[!h]
    \centering
    \begin{tabular}{cc}
        \hline
        Variable & Value and description \\
        \hline
        TA Details & Variance only, no pooling \\
        Power Model & HW model \\
        \# profiling traces & 50\,000 \\
        \# attack traces & 300 \\
        Correction Factor & 10 \\
        Population size & 50 \\
        \# Iterations & 20 \\
         \hline
    \end{tabular}
    \caption{Attack specifications (Unprotected AES)}
    \label{uc2.details}
\end{table}

\begin{figure}[!h]
\begin{minipage}{0.47\textwidth}
\centering
\begin{tabular}{cccc}
\hline
 & Factor name & Low setting & High setting\\
\hline
A &  CF &  1 & 10 \\
B &  \# Iter. & 10 & 20 \\
C &  Pop. Size & 20 & 50 \\
\hline
\end{tabular}
\captionof{table}{Working variables and values (DoE)}
\label{ParameterSelectionASCAD}
\end{minipage}\hfill
\begin{minipage}{0.47\textwidth}
\centering
\includegraphics[width=1\textwidth]{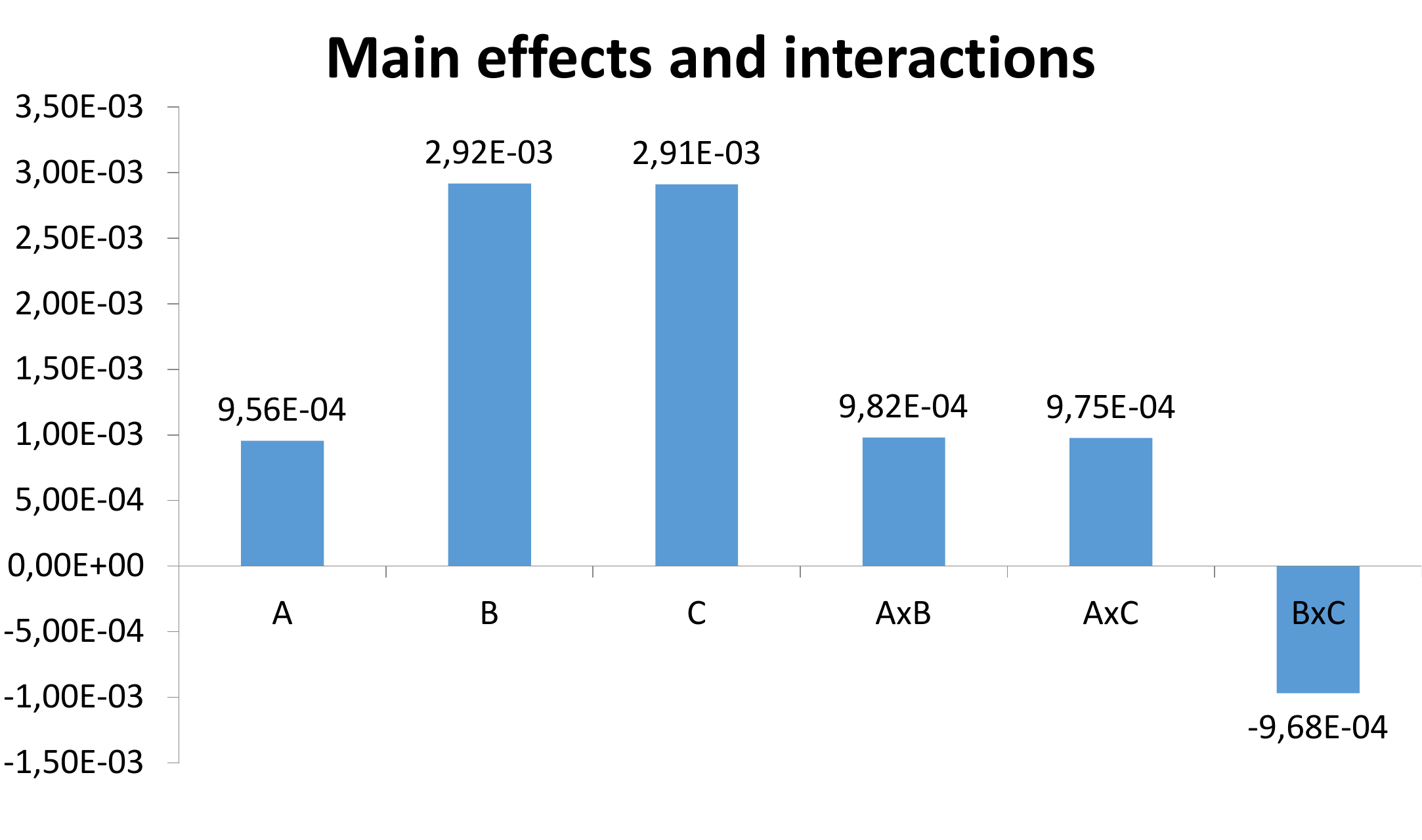}
\captionof{figure}{\label{DOE_ASCAD} Effects and Iterations (DoE)}
\end{minipage}\hfill
\vspace{0.5cm}
\begin{minipage}{1\textwidth}
\centering
\begin{tabular}{ccccccc}
\hline
Exp & A & B & C & $ n_{POI} $ & $ge$ & \textbf{Eval}\\
\hline
1 & 1 & 10 & 20 & 51 & 2 & -0.007813\\
2 & 1 & 10 & 50 & 69 & 2 & -0.007813\\
3 & 1 & 20 & 20 & 72 & 2 & -0.007813\\
4 & 1 & 20 & 50 & 42 & 1 & -6.56E-05\\

5 & 10 & 10 & 20 & 71 & 4 & -0.156250\\
6 & 10 & 10 & 50 & 66 & 1 & -0.001031\\
7 & 10 & 20 & 20 & 50 & 1 & -7.81E-04\\
\textbf{8} & \textbf{10} & \textbf{20} & \textbf{50} & \textbf{32} & \textbf{1} & \textbf{-5.00E-04}\\
\hline
\end{tabular}
\captionof{table}{Table of results (DoE)}
\label{ExperimentRoundsASCAD}
\end{minipage}\hfill
\end{figure}

Again, we need to adjust the correction factor, the number of iterations, and the population size (Table \ref{ParameterSelectionASCAD}). In order to do so, another experimental design has been performed (Fig.~\ref{DOE_ASCAD} and Tables~\ref{ParameterSelectionASCAD} and~\ref{ExperimentRoundsASCAD}.). Here we have selected different low and high settings for the variables B and C (higher values, Table~\ref{ParameterSelectionASCAD}).  This is because this use case is considerably more difficult than the previous one as we are targeting a protected AES implementation and we can not accelerate the search using a POI selection graphic. 


In practice, this would commonly imply having a wider population size and more iterations, and hence we consider bigger values for the experimental design (Table~\ref{ParameterSelectionASCAD}). In fact, the outcome of the experimental design confirms that we obtain better results with the setup used in Experiment 8 (correction factor of 10, 20 iterations, and population size of 50), although we succeed in all the 8 experiments of the DoE anyway. 




It should be noticed that even bigger values for variables A, B, and C could have been selected. However, since we succeed in all the 8 experiments, we claim that it is not necessary because the improvement is not very significant considering the increase in computational effort required. Therefore, as our experimental design shows, most of the combinations show state-of-the-art performance, which is another strength of this methodology. Nevertheless, we select the setup in Experiment 8 for the experiments as it shows a good balance between performance and resources.

\subsubsection{Results\\}

\begin{figure}[!h]
    \centering
    \includegraphics[width=1\textwidth]{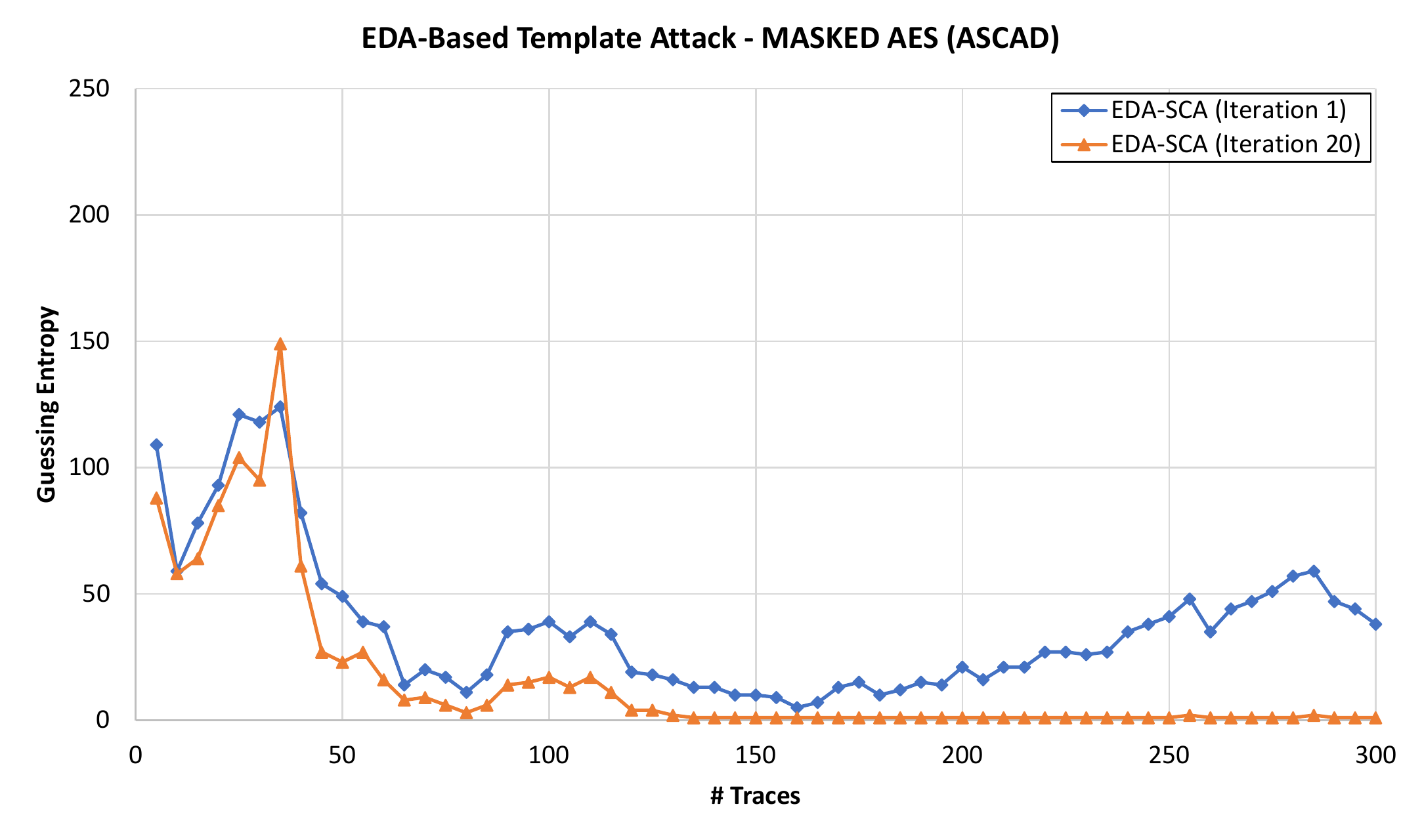}
    \caption{Results on ASCAD dataset (Guessing Entropy)} 
    \label{fig:geASCAD}
\end{figure}

In Tables \ref{uc2.ITER1} and \ref{uc2.ITER20}, results from the first and last iteration are shown. Please note that they are shown in the same manner as in the previous use case. We can observe that although in the first iteration the results are quite poor, in the last iteration we are able to succeed in the attack with all the 50 individuals. In addition, a graphical representation of the results can be observed in Fig.~\ref{fig:geASCAD}. 

\begin{table}[!h]
\begin{minipage}{0.47\textwidth}
    \centering
	\begin{tabular}{cccc}
	\hline
	Ind & Eval & $ n_{POI} $ & $ge$ \\
	\hline
    1 & -1.484375 & 76 & 38 \\
    2 & -1.679688 & 74 & 43 \\
    3 & -1.875000 & 60 & 48 \\
    4 & -1.914063 & 81 & 49 \\
    5 & -2.031250 & 64 & 52 \\
    6 & -2.304688 & 71 & 59 \\
    7 & -2.421875 & 81 & 62 \\
    8 & -2.460938 & 70 & 63 \\
    9 & -2.500000 & 58 & 64 \\
    10 & -2.656250 & 64 & 68 \\
    ···  & ···  & ···  & ··· \\
    48 & -4.687500 & 71 & 120 \\
    49 & -5.117188 & 63 & 131 \\
    50 & -6.914063 & 59 & 177 \\
    \hline
	\end{tabular}
	\caption{Results of the first iteration}
	\label{uc2.ITER1}
\end{minipage}\hfill
\begin{minipage}{0.47\textwidth}
    \centering
	\begin{tabular}{cccc}
	\hline
	Ind & Eval & $ n_{POI} $ & $ge$ \\
	\hline
    1 & -5.00E-04 & 32 & 1 \\
    2 & -5.16E-04 & 33 & 1 \\
    3 & -5.31E-04 & 34 & 1 \\
    4 & -5.31E-04 & 34 & 1 \\
    5 & -5.31E-04 & 34 & 1 \\
    6 & -5.31E-04 & 34 & 1 \\
    7 & -5.31E-04 & 34 & 1 \\
    8 & -5.47E-04 & 35 & 1 \\
    9 & -5.47E-04 & 35 & 1 \\
    10 & -5.47E-04 & 35 & 1 \\
        ···  & ···  & ···  & ··· \\
    48 & -7.19E-04 & 46 & 1 \\
    49 & -7.34E-04 & 47 & 1 \\
    50 & -7.81E-02 & 40 & 2 \\
    \hline
	\end{tabular}
	\caption{Results of the last iteration}
	\label{uc2.ITER20}
\end{minipage}\hfill  
\end{table}

On the other hand, if we compare these results with related works, it is clear from the facts that our EDA-Based PA approach can provide state-of-the-art results (see Table~\ref{uc2.comparison}). There we can observe the amount of attack traces required to reach Guessing Entropy 0 ($\bar{N} t_{G E}$) from the work in \cite{prouff2018ascad} using TAs, the work in \cite{zaid2019methodology} which is, to the best of our knowledge, the deep learning model which better performance has in this dataset, and our approach.  It should be noticed that a meaningful comparison should require also comparing computation resources. However, as our technique is at a very early stage and there is still a lot of room for optimizing its computation, and in \cite{zaid2019methodology} they do not mention their spent computation resources, we just compare our results in terms of guessing entropy.

\begin{table}[!h]
    \centering
    \begin{tabular}{cDDD}  
        \hline
         & Template Attacks \cite{prouff2018ascad} & Deep Learning
         \cite{zaid2019methodology} & EDA-Based PA\\
         \hline
         $\bar{N} t_{G E}$  & $\approx450$ & $191$ & $\approx150$\\
        \hline
    \end{tabular}
    \caption{Comparison of performance on ASCAD}
    \label{uc2.comparison}
\end{table}

In conclusion, our approach has better performance, in terms of guessing entropy, than the attacks using Deep Learning performed by Zaid \textit{et al.} in \cite{zaid2019methodology} and TAs performed by Prouff \textit{et al.} in \cite{prouff2018ascad}.

\section{Introduction of the \textit{AES\_PT} dataset}
\label{introAESPT}

Motivated by the lack of an open dataset for Side-channel analysis which includes traces from different copies of the same device\footnote{Note that there exist the \textit{Grizzly} dataset \cite{Choundary2018efficient}, but it includes power-analysis traces of four copies of the same device performing an 8-bit load instruction, not a cryptographic implementation.} we have generated the \textit{AES\_PT} datase\footnote{AES\_PT Dataset: \url{https://github.com/urioja/AESPT}} (PT stands for portable). It includes power analysis traces from four copies of the same development board mounting an STM32F411VE~\cite{UCdatasheet} high-performance Arm\textsuperscript{\tiny\textregistered} Cortex\textsuperscript{\tiny\textregistered} -M4 32-bit RISC microcontroller working at 100 MHz. 
The four copies are called D1, D2, D3, and D4 from now on. As mentioned in the introduction, this dataset was created with the idea of making ``realistic'' TAs and therefore includes subsets of traces of each clone device performing unprotected and protected AES-128 implementations (see Sec.~\ref{AESimpl}), with both fixed and random cryptographic keys. 

\subsection{Acquisition specifications}

\begin{figure}[!h]
    \centering
    \includegraphics[width=0.8\textwidth]{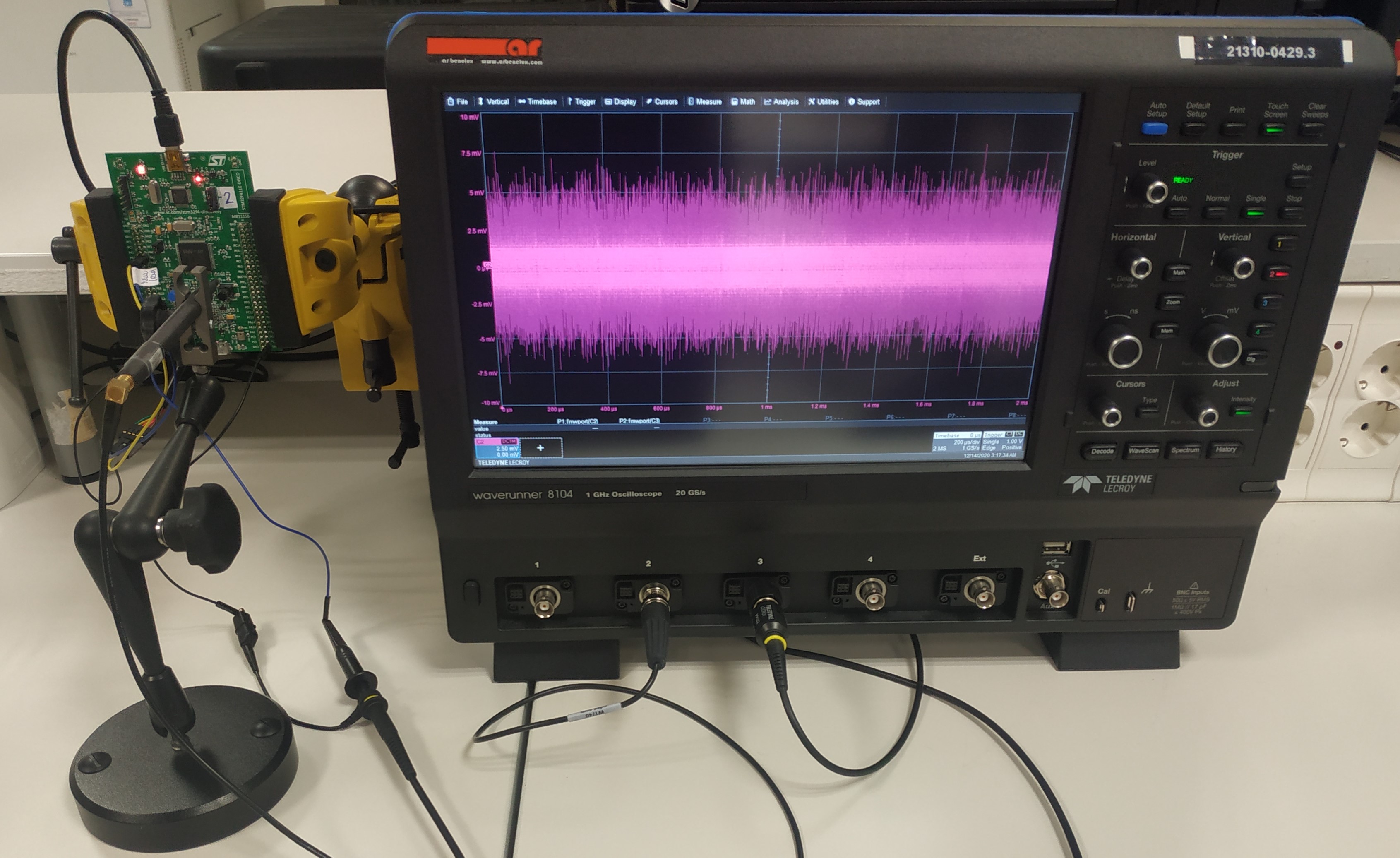}
    \caption{Experimental setup}
    \label{fig:Exp_Setup}
\end{figure}

Fig.~\ref{fig:Exp_Setup} shows a picture of the experimental setup used to acquire the power traces. In more detail, the devices are encrypting 16-byte random plaintexts using two software AES implementations: unprotected AES and masked AES (see Sec.~\ref{AESimpl}). During that operation, we measure the power consumption of the device with a Langer EM probe attached to a 20 GS/s digital oscilloscope (LeCroy Waverunner 9104) triggered by the microcontroller, which rises a GPIO signal when the internal computation starts. The high sensibility probe is placed over a decoupling capacitor connected to the power line of the device. Each power trace is formed by 1225 samples (2300 for the masked implementation) taken at 1GHz with 8-bit resolution, corresponding to the first Sbox operation. Traces are preprocessed by applying zero mean, standardization, waveform realignment, and a lightweight software lowpass filter. Nevertheless, traces are deliberately quite noisy (due to the nature of EM measurements, variations during the acquisition of the traces, constructive differences between the devices, etc.) to serve to represent realistic experimental use cases. 

\subsection{AES Implementation}
\label{AESimpl}

As mentioned above, the \textit{AES\_PT} dataset includes traces of each clone device performing both unprotected AES and masked AES. Both algorithms are implemented in C language. The unprotected AES implementation is a regular AES-128 (in ECB mode) software implementation\footnote{Small portable AES-128 in C: \url{https://github.com/bitdust/tiny-AES128-C}}. Regarding the masked implementation, it is a modification of the previous one which matches the same masking method described in~\cite{BlueBook} (Masked Lookup Table). Below there is a brief explanation of this masking method.\\

\noindent
\textbf{Masked Lookup Table:} In a masked implementation, each intermediate value $ v $ is concealed by a random value $ m $ that is called mask, which is different for each execution and unknown by the attacker, such that $ v_{m} = v * m $. This randomizes the intermediate value, mitigating the dependency between the power consumption of the DUT and the processed intermediate value, and providing security against firt-order DPA attacks. 

In SCA over software AES implementations on microcontroller, it is common to target the Sbox output as sensitive intermediate value $ v =\operatorname{Sbox}[p \oplus k] $. The Sbox operation is a substitution box (lookup table) used in the Rijndael cipher~\cite{daemen01,daemen02}. Since this dataset includes traces corresponding to the first Sbox operation only, we will focus on that part of the masking method (see~\cite{BlueBook} for the full explanation, including the masking of the rest of intermediate values). 

Following the approach in~\cite{BlueBook}, the Sbox operation is masked using two masks: the input mask $m$ and the output mask $m'$. At the beginning of each AES encryption, a masked Sbox table $S_{m}$ is computed with the property $S_{m}(x \oplus m) = S(x) \oplus m $, and used instead of the original table. Generating the masked table is a simple process, as one only has to run through all inputs $ x $, look up $S(x)$ and store $S(x) \oplus m$ (Alg. \ref{alg:Mask}). Conversely, it should be noticed that it increases the computational effort and the amount of memory used by the microcontroller.

\begin{algorithm}[h]
\caption{Mask Sbox function}
\label{alg:Mask}
\SetKwInput{KwInput}{Input}                
\SetKwInput{KwOutput}{Output}              
\DontPrintSemicolon

\KwInput{a 256-byte unmasked Sbox  ($S[1],\ldots,S[256]$),}
\myinput{an 1 Byte input mask ($m\_in$),}
\myinput{an 1 Byte output mask ($m\_out$)}
\KwOutput{a 256-byte masked Sbox  ($Sm[1],\ldots,Sm[256]$),}
\SetKwFunction{FMain}{mask\_Sbox}
 
\SetKwProg{Fn}{Function}{:}{}
\Fn{\FMain{S, m\_in, m\_out}}{
\For{$i\gets0$ \KwTo $256$ \KwBy $1$}{
Sm[i^m\_in] = S[i]^m\_out;
}
\KwRet Sm\;
}
\end{algorithm}

\subsection{Dataset organization}

\begin{figure}[!t]
    \centering
    \includegraphics[width=1\textwidth]{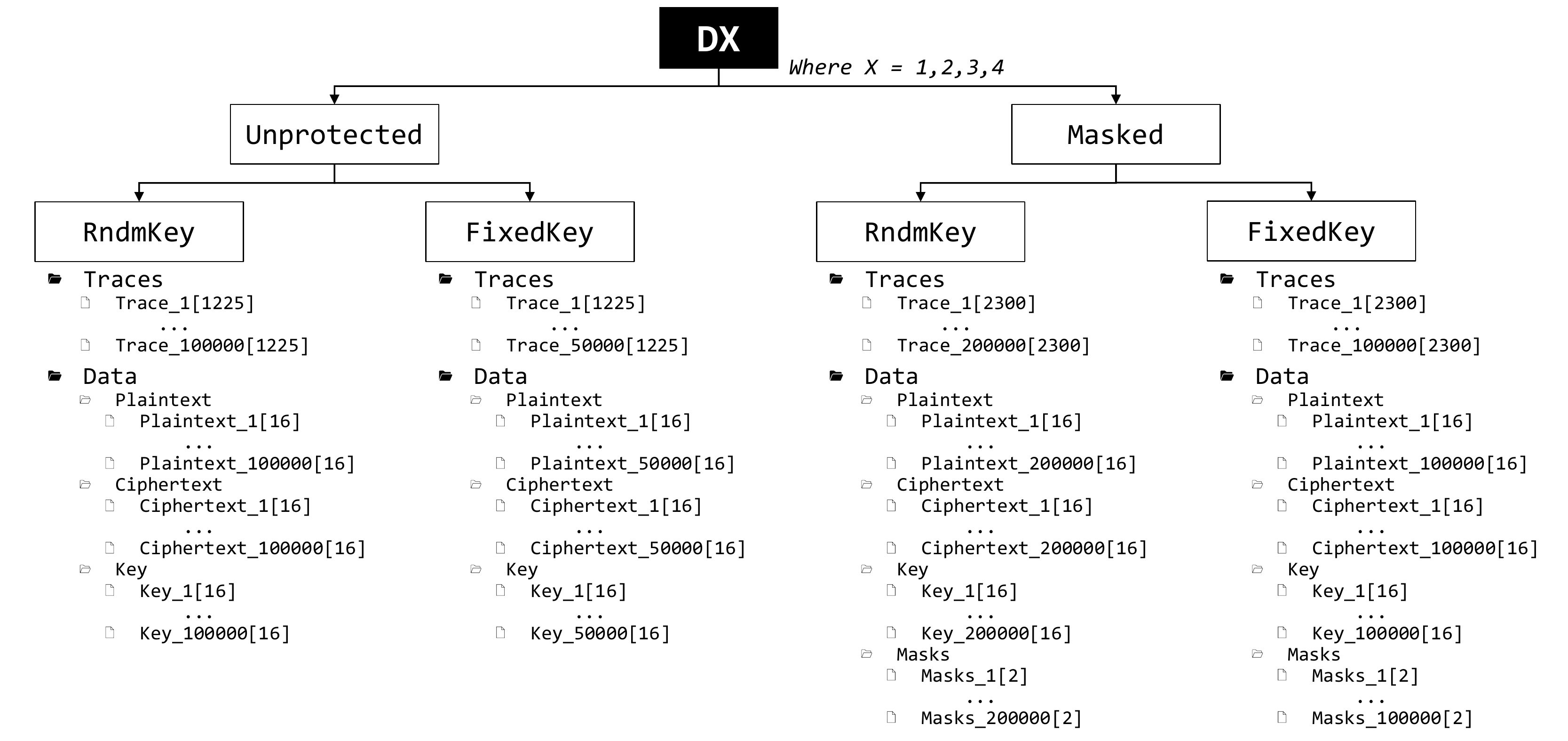}
    \caption{\textit{AES\_PT} dataset organization (per device)}
    \label{fig:AES_PT}
\end{figure}

The dataset is stored using the HDF5 format. The entire dataset is contained in a single file \textit{AES\_PT.h5} which has 4 groups, one per each clone device, called D1, D2, D3, and D4. In turn, each group is divided into smaller subgroups, as it includes traces of the device performing unprotected and masked AES implementations with both fixed key and random key. An organization chart of how each device group is structured can be seen in Fig.~\ref{fig:AES_PT}. In short, the dataset includes 600\,000 traces per device:
\begin{itemize}
    \item 150\,000 unprotected AES power traces (100\,000 traces of the device using random keys and 50\,000 traces of the device using a fixed key) 
    \item 300\,000 masked AES power traces (200\,000 traces of the device using random keys and 100\,000 traces of the device using a fixed key)
\end{itemize}

In turn, each set of traces includes its corresponding associated data: plaintext, ciphertext, key and mask (input and output mask, only for the Masked AES implementations).

\section{Experimental use cases: Several devices (considering portability)}
\label{sec.EUCport}
In this section, we repeat the previous use cases but considering portability. This is accomplished by using the aforementioned \textit{AES\_PT} dataset.
    
\subsection{Unprotected AES implementation on microcontrollers (\textit{AES\_PT})}
\label{UC3}

\subsubsection{Target description\\}
Our targets are four copies of the same development board mounting an STM32F411VE, as explained in the previous section. As an attacker, our goal is to guess the secret key used to encrypt data. A set of $ n_p = 10000 $ profiling traces are taken from the profiling device (D1) using random keys and plaintext. In the \textit{attack phase} a set of  $ n_a = 100 $ traces of D1, D2, D3, and D4 encrypting random plaintext with a fixed key (unknown by the attacker) are taken. Then the multivariate model is applied and the secret key is guessed using the maximum likelihood principle. We are using reduced templates (variance only) without compression method (no pooling). As the sensitive intermediate value we use the Hamming Weight of an Sbox output: 

\begin{equation}
Y^{(i)}\left(k^{*}\right)=\operatorname{Sbox}[P_{0}^{(i)} \oplus k^{*}]  
\end{equation}

\subsubsection{Attack details\\}
\label{UC3.ad}

\begin{table}[!h]
    \centering
    \begin{tabular}{cc}
        \hline
        Variable & Value and description \\
        \hline
        TA Details & Variance only, no pooling \\
        Power Model & HW model \\
        \# profiling traces & 10\,000 \\
        \# attack traces & 100 \\
        Correction Factor & 10 \\
        Population size & 50 \\
        \# Iterations & 20 \\
         \hline
    \end{tabular}
    \caption{Attack specifications (Unprotected AES)}
    \label{uc3.details}
\end{table}

Table \ref{uc3.details} summarizes the specifications of the attack. Although we could have performed another experimental design, this time we have directly selected a correction factor of 10, a population size of 50 and 30 iterations (based on the results of previous experiments) in order to save time. As evaluation function, we consider the following formula to evaluate each individual (POI candidate):

\begin{equation}
\label{eq.uc3}
Eval(ind) = \left \{ \begin{matrix} - CF*ge_{All} & \mbox{if }ge_{All}\neq\mbox{1}
\\ - CF*\frac{n_{POI}}{n_{samples}}*ge_{All} & \mbox{if }ge_{All}=\mbox{ 1}\end{matrix}\right.   
\end{equation}

\begin{figure}[!h]
    \centering
    \includegraphics[width=0.7\textwidth]{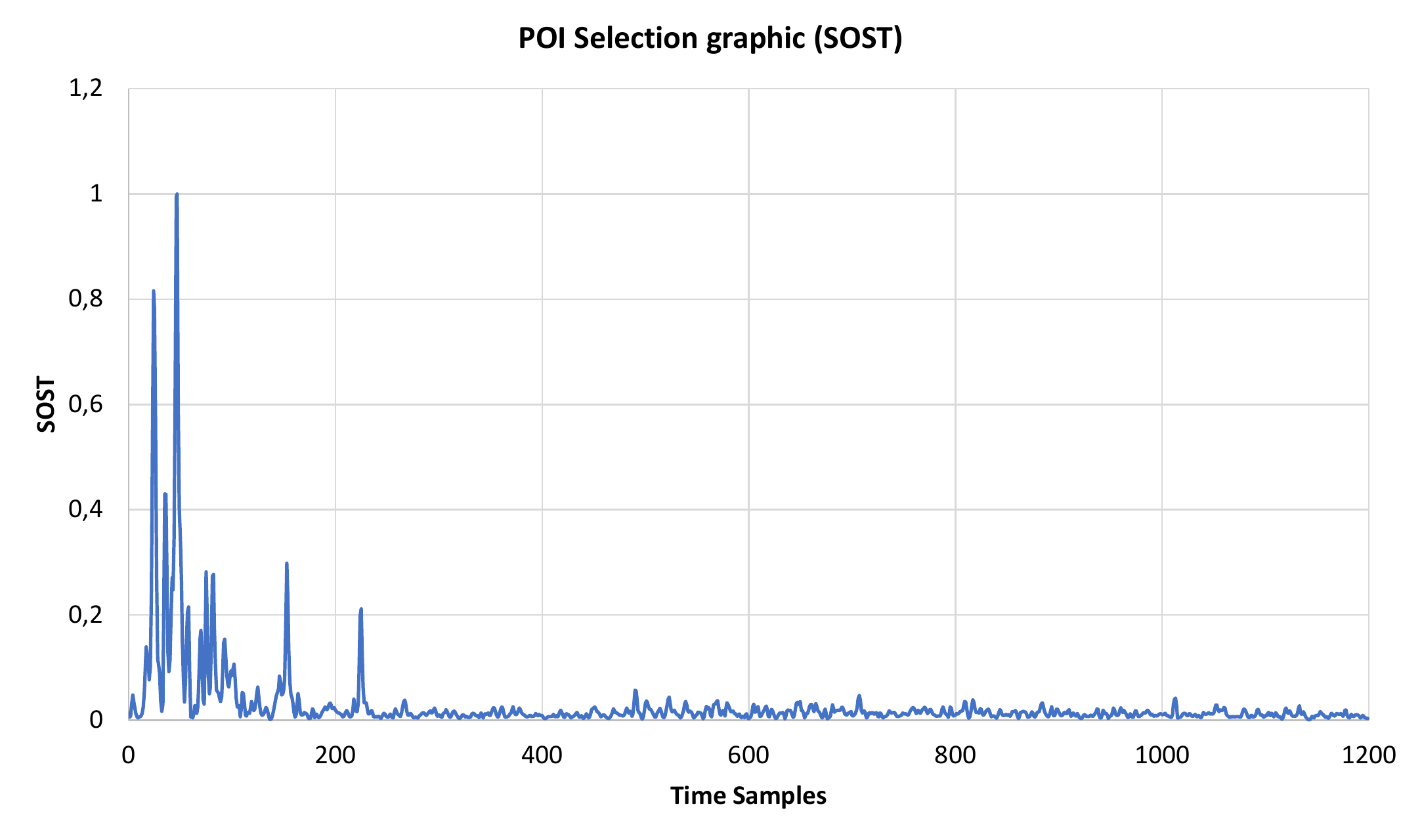}
    \caption{POI selection graphic (SOST) of the targeted intermediate value (1st Byte Sbox output), normalized between 0 and 1}
    \label{FSG_PLOT}
\end{figure}

Where $ge_{All}$ is the sum of the attack results ($ge$) of attacking devices D1, D2, D3 and D4 respectively ($ge_{All} = \frac{ge_{D1}}{256} + \frac{ge_{D2}}{256} + \frac{ge_{D3}}{256} + \frac{ge_{D4}}{256}$) using the model built from D1\footnote{We divide $ge$ by 256 (maximum guessing entropy in this case) in order to bound the value between 0 and 1}. Please note that it is the same evaluation function as in previous use cases, except for $ge_{All}$. In addition, we are using a POI selection graphic (Fig. \ref{FSG_PLOT}) obtained from the profiling traces to guide the search even more (as explained in Sec. \ref{UC1}).

\subsubsection{Results\\}

In Tables \ref{uc3.ITER1} and \ref{uc3.ITER20}, results from the first and last iteration are shown. Please note that they are shown in the same manner as in the previous use cases. In this case, since we are considering portability but the implementation is weak (unprotected implementation), our technique throws good results from the beginning as some individuals of the first iteration show a relatively good performance. Nevertheless, after 20 iterations we are able to succeed in the attack in all copies with all the 50 individuals of the population, a significantly better performance.

\begin{table}[!h]
\begin{minipage}{0.47\textwidth}
    \centering
	\begin{tabular}{ccccccc}
	\hline
	Ind & Eval & $ n_{POI} $ & $ge_{D1}$ & $ge_{D2}$ & $ge_{D3}$ & $ge_{D4}$ \\
	\hline
    1 & -0.000010 & 15 & 1 & 18 & 22 & 11 \\
    2 & -0.000056 & 14 & 10 & 1 & 16 & 151 \\
    3 & -0.000143 & 14 & 30 & 38 & 18 & 3 \\
    4 & -0.000302 & 11 & 6 & 75 & 18 & 16 \\
    5 & -0.000418 & 14 & 11 & 33 & 15 & 33 \\
    6 & -0.000440 & 17 & 3 & 106 & 3 & 198 \\
    7 & -0.000827 & 19 & 2 & 137 & 16 & 81 \\
    8 & -0.001223 & 15 & 6 & 178 & 2 & 246 \\
    9 & -0.001321 & 10 & 11 & 63 & 13 & 63 \\
    10 & -0.001747 & 13 & 20 & 59 & 6 & 106 \\
        ···  & ···  & ···  & ···  & ···  & ···  & ···\\
    48 & -0.637708 & 14 & 116 & 198 & 75 & 159 \\
    49 & -0.655694 & 21 & 132 & 232 & 44 & 209 \\
    50 & -1.337067 & 14 & 113 & 231 & 88 & 250 \\
    \hline
	\end{tabular}
	\caption{Results of the first iteration}
	\label{uc3.ITER1}
\end{minipage}\hfill
\begin{minipage}{0.47\textwidth}
    \centering
	\begin{tabular}{ccccccc}
	\hline
	Ind & Eval & $ n_{POI} $ & $ge_{D1}$ & $ge_{D2}$ & $ge_{D3}$ & $ge_{D4}$ \\
	\hline
    1 & -1.58E-11 & 17 & 1 & 1 & 1 & 1 \\
    2 & -1.58E-11 & 17 & 1 & 1 & 1 & 1 \\
    3 & -1.68E-11 & 18 & 1 & 1 & 1 & 1 \\
    4 & -1.68E-11 & 18 & 1 & 1 & 1 & 1 \\
    5 & -1.68E-11 & 18 & 1 & 1 & 1 & 1 \\
    6 & -1.77E-11 & 19 & 1 & 1 & 1 & 1 \\
    7 & -1.77E-11 & 19 & 1 & 1 & 1 & 1 \\
    8 & -1.86E-11 & 20 & 1 & 1 & 1 & 1 \\
    9 & -1.86E-11 & 20 & 1 & 1 & 1 & 1 \\
    10 & -1.86E-11 & 20 & 1 & 1 & 1 & 1 \\
        ···  & ···  & ···  & ···  & ···  & ···  & ···\\
    48 & -6.98E-09 & 16 & 1 & 3 & 1 & 1 \\
    49 & -1.16E-08 & 16 & 1 & 5 & 1 & 1 \\
    50 & -1.40E-08 & 18 & 2 & 3 & 1 & 1 \\
    \hline
	\end{tabular}
	\caption{Results of the last iteration}
	\label{uc3.ITER20}
\end{minipage}\hfill  
\end{table}

\begin{figure}[hbtp]
    \centering
    \includegraphics[width=1\textwidth]{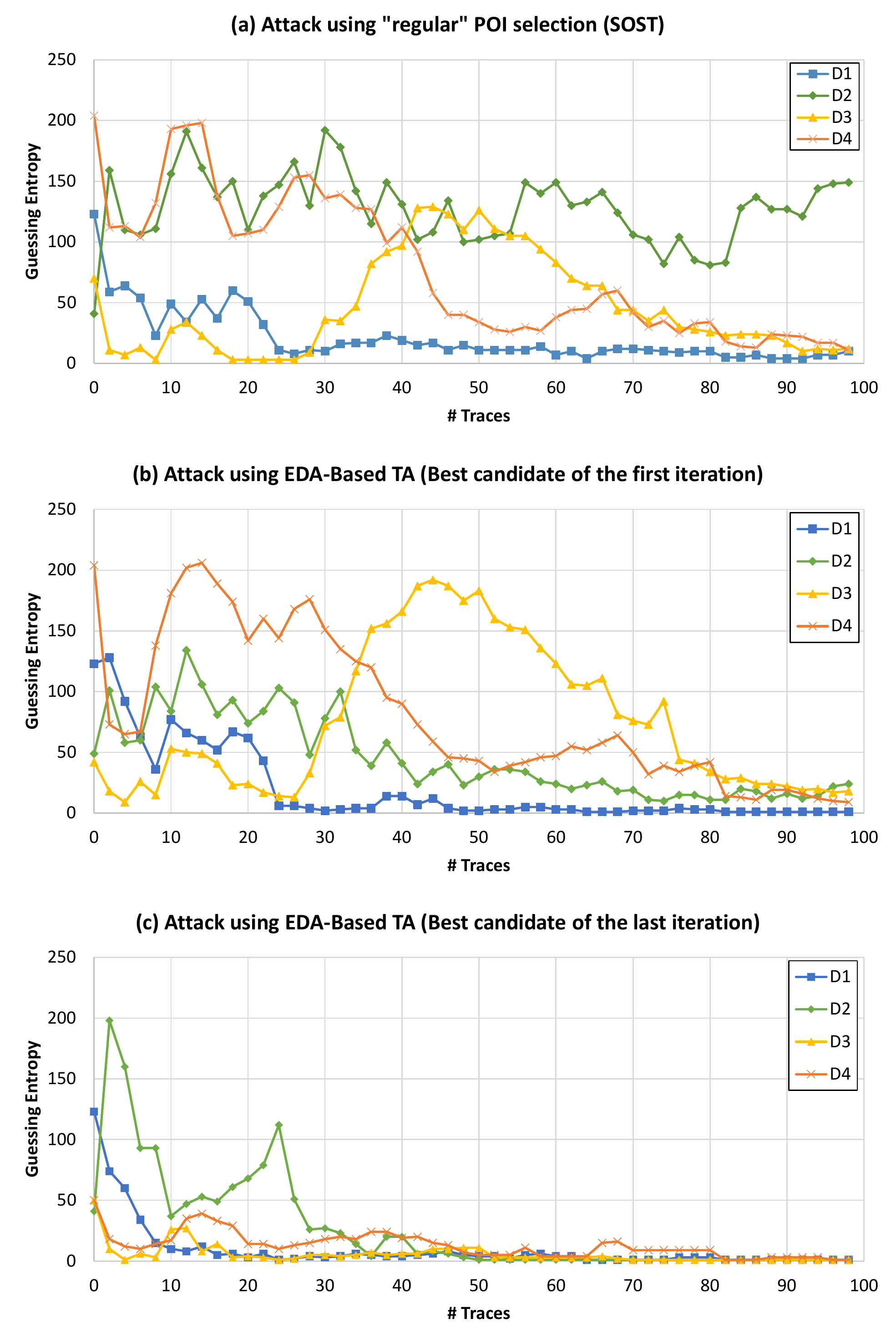}
    \caption{Results on \textit{AES\_PT} dataset (unprotected AES)} 
    \label{fig:gePlotsDISCOunprotectedAES}
\end{figure}

A more graphical representation of the obtained results can be observed in Fig. \ref{fig:gePlotsDISCOunprotectedAES}. There you can see the experimental results (guessing entropy) of (a) using a ``regular'' POI selection (just select 20 POI on the highest SOST values), (b) the results of the best individual (POI selection candidate) of the first iteration and (c) the results of the best individual (POI selection candidate) of the last iteration. It should be noticed that the ``regular'' POI selection (a) throws the worse results: Although the performance of the attacks over D3 and D4 are not so bad, the attack over D2 is completely inconclusive (the model generated with D1 does not represent clearly the leakage of D2). Conversely, the best candidate of both the first and last iteration (Fig. \ref{fig:gePlotsDISCOunprotectedAES} (b) and (c)) have better performance. Even though the results in (b) do not show a huge improvement against (a) (Actually, we achieve a similar performance except for D2), the results of the last iteration are far better than the previous one (The rank of the correct key candidate is 1 for all copies after 100 traces). In conclusion, these results demonstrate the suitability of this method when a portable template attack scenario is considered.

\subsection{Protected AES implementation}
\label{UC4}

\subsubsection{Target description\\}
In this use case, our targets are the same as in the previous use case except that we are using a protected AES implementation (the masked countermeasure described in \cite{BlueBook}).

\subsubsection{Attack Details\\}
\label{UC4.ad}

Table \ref{uc4.details} summarizes the specifications of the attack. Again, although we could have performed another experimental design, this time we have directly selected a correction factor of 10, a population size of 50 and 30 iterations (based on the results of previous experiments) in order to save time. As evaluation function, we consider the same formula as in the previous use case (Eq. \ref{eq.uc3}) to evaluate each individual (POI candidate).

\begin{table}[!h]
    \centering
    \begin{tabular}{cc}
        \hline
        Variable & Value and description \\
        \hline
        TA Details & Variance only, no pooling \\
        Power Model & HW model \\
        \# profiling traces & 10\,000 \\
        \# attack traces & 500 \\
        Correction Factor & 10 \\
        Population size & 50 \\
        \# Iterations & 20 \\
         \hline
    \end{tabular}
    \caption{Attack specifications (Protected AES)}
    \label{uc4.details}
\end{table}

\subsubsection{Results\\}

\begin{figure}[!h]
    \centering
    \includegraphics[width=1\textwidth]{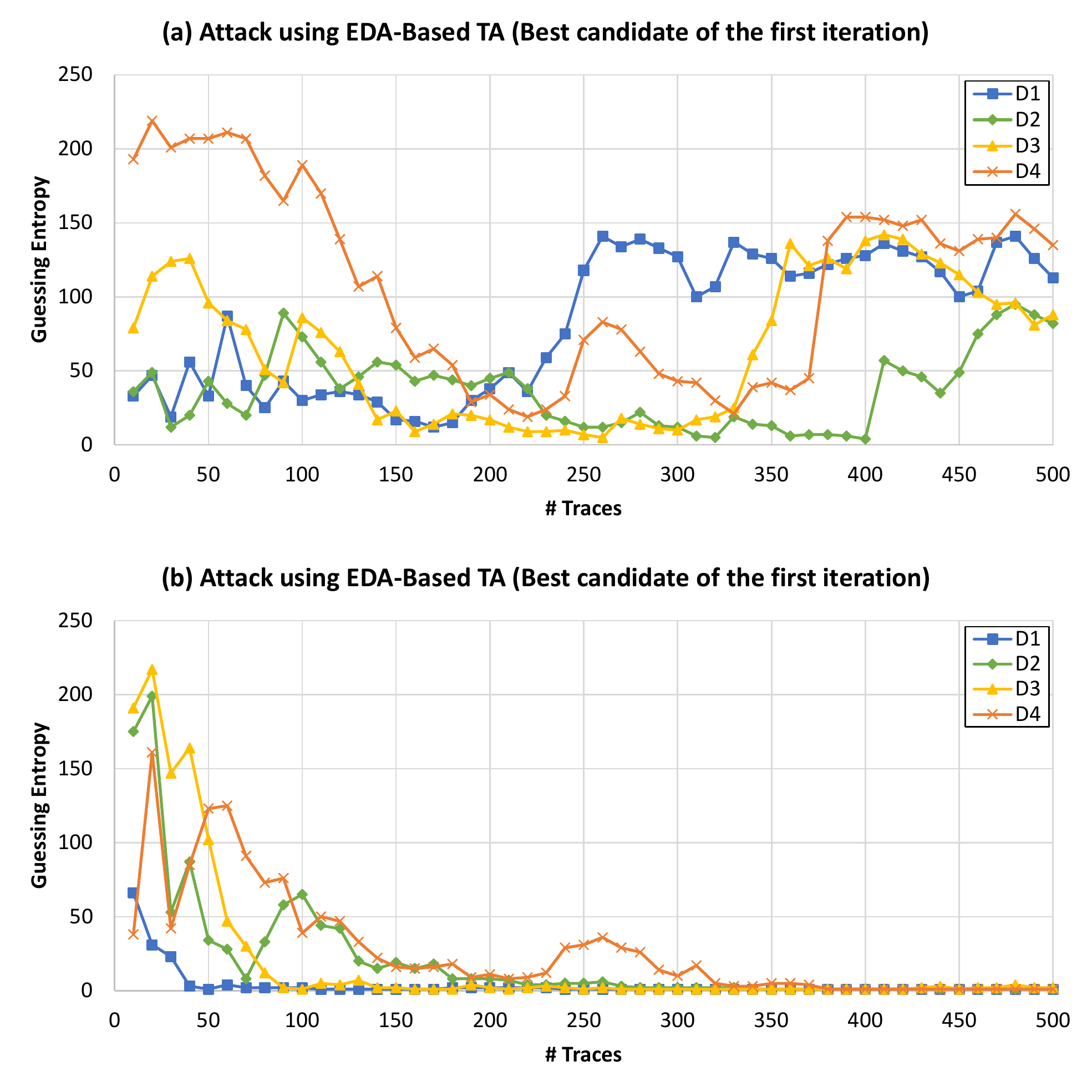}
    \caption{Results on \textit{AES\_PT} dataset (Masked AES)} 
    \label{fig:geUC4}
\end{figure}

In Tables \ref{uc4.ITER1} and \ref{uc4.ITER30}, results from the first and last iteration are shown. Please note that they are shown in the same manner as in the previous use case, but we have not represented some of the individuals in order to reduce the tables' size. A graphical representation of the results can be observed in Fig. \ref{fig:geUC4}. We can observe how although in the first iteration the results are quite weak, in the last iteration we are able to succeed in the attack on the four copies with a model built from D1, obtaining similar results to our previous use case even when the AES implementation is protected using Masking. This demonstrates the suitability of EDA-Based PA in a portable template attack scenario, even with countermeasures and noisy traces.

\begin{table}[!h]
\begin{minipage}{0.47\textwidth}
    \centering
	\begin{tabular}{ccccccc}
	\hline
	Ind & Eval & $ n_{POI} $ & $ge_{D1}$ & $ge_{D2}$ & $ge_{D3}$ & $ge_{D4}$ \\
	\hline
    1 & -2.56E-01 & 129 & 113 & 82 & 88 & 135 \\
    2 & -3.13E-01 & 114 & 150 & 137 & 252 & 26 \\
    3 & -5.71E-01 & 111 & 71 & 165 & 249 & 84 \\
    4 & -5.84E-01 & 104 & 124 & 176 & 221 & 52 \\
    5 & -7.70E-01 & 111 & 60 & 243 & 244 & 93 \\
    6 & -9.53E-01 & 112 & 73 & 177 & 214 & 148 \\
    7 & -9.76E-01 & 98 & 119 & 88 & 188 & 213 \\
    8 & -1.039606 & 101 & 134 & 76 & 226 & 194 \\
    9 & -1.068624 & 105 & 72 & 204 & 144 & 217 \\
    10 & -1.089144 & 110 & 118 & 211 & 244 & 77 \\
    ···  & ···  & ···  & ···  & ···  & ···  & ···\\
    48 & -4.566987 & 110 & 167 & 245 & 251 & 191 \\
    49 & -5.257555 & 112 & 228 & 181 & 251 & 218 \\
    50 & -5.661285 & 106 & 198 & 205 & 256 & 234 \\
    \hline
	\end{tabular}
	\caption{Results of the first iteration}
	\label{uc4.ITER1}
\end{minipage}\hfill
\begin{minipage}{0.47\textwidth}
    \centering
	\begin{tabular}{ccccccc}
	\hline
	Ind & Eval & $ n_{POI} $ & $ge_{D1}$ & $ge_{D2}$ & $ge_{D3}$ & $ge_{D4}$ \\
	\hline
    1 & -4.66E-09 & 97 & 1 & 1 & 2 & 1 \\
    2 & -4.66E-09 & 90 & 1 & 1 & 2 & 1 \\
    3 & -4.66E-09 & 95 & 1 & 1 & 2 & 1 \\
    4 & -6.98E-09 & 96 & 1 & 1 & 3 & 1 \\
    5 & -6.98E-09 & 102 & 1 & 1 & 3 & 1 \\
    6 & -9.31E-09 & 92 & 1 & 1 & 4 & 1 \\
    7 & -9.31E-09 & 93 & 1 & 1 & 4 & 1 \\
    8 & -9.31E-09 & 97 & 1 & 1 & 4 & 1 \\
    9 & -9.31E-09 & 95 & 1 & 1 & 4 & 1 \\
    10 & -9.31E-09 & 100 & 1 & 1 & 4 & 1 \\
    ···  & ···  & ···  & ···  & ···  & ···  & ···\\
    48 & -2.79E-08 & 99 & 1 & 1 & 12 & 1 \\
    49 & -3.26E-08 & 108 & 1 & 1 & 14 & 1 \\
    50 & -5.12E-08 & 99 & 1 & 1 & 22 & 1 \\
    \hline
	\end{tabular}
	\caption{Results of the last iteration}
	\label{uc4.ITER30}
\end{minipage}\hfill  
\end{table}

\section{Conclusions and future works}
\label{sec.concl}

On the whole, our experimental use cases demonstrate the suitability of our method for automated TA optimization in the context of AES implementations on small embedded devices, for different implementations and leakage models. Nevertheless, this approach is not algorithm-dependent so we claim that this approach can be applied to other scenarios that may be considered in future lines of work. Moreover, although we have used TA for demonstrating our approach, other kinds of PA are potential candidates to be performed in combination with EDAs.

Our EDA-Based PA is able to obtain state-of-the-art results in a relatively straightforward way: it can heuristically find the POIs with better performance automatically and efficiently. Moreover, we have shown how even in the most adverse use case (masked AES implementation considering portability), our technique provides particularly good results as we can attack several devices with a leakage model built only from one copy.

In addition, we claim that our approach mitigates the need for a human on the loop: as the attacks are performed automatically, this approach may help technicians without a deep knowledge of all the basics involved in TAs to perform this part of the evaluation process properly. Besides, this approach could be also interesting to experts in evaluation laboratories since it allows them to parallelize tasks and reduce the time cost of the evaluation process.

The method at the moment works in a matter of hours but, as future works, we have identified many ways of optimizing the method, e.g. attack parallelization, computation optimization, etc. which could reduce drastically the computation time.

%
%
%
\bibliographystyle{splncs04}
\bibliography{bib.bib}

\begin{thebibliography}{10}
\providecommand{\url}[1]{\texttt{#1}}
\providecommand{\urlprefix}{URL }
\providecommand{\doi}[1]{https://doi.org/#1}

\bibitem{BSI}
Federal {O}ffice for {I}nformation {S}ecurity ({BSI}) - {C}ommon {C}riteria for
  examination and evaluation of it security,
  \url{https://www.bsi.bund.de/EN/Topics/CommonCriteria/commoncriteria.html},
  last accessed: December 31, 2020

\bibitem{DPAcontest}
Telecom paristech sen research group: Dpa contest (4thedition) (2013–2014),
  \url{http://www.DPAcontest.org/v4/.}

\bibitem{ANSSI}
{A}gence {N}ationale de la {S}écurité des {S}ystèmes d'information -
  {C}ertified products (2019),
  \url{https://www.ssi.gouv.fr/en/products/certified-products/}, last accessed:
  December 31, 2020

\bibitem{daemen01}
197, F.I.P.S.P.: {A}nnouncing the {A}dvanced {E}ncryption {S}tandard ({AES})
  (nov 2001), \url{https://nvlpubs.nist.gov/nistpubs/FIPS/NIST.FIPS.197.pdf}

\bibitem{Archambeau2006subspaces}
Archambeau, C., Peeters, E., Standaert, F.X., Quisquater, J.J.: Template
  attacks in principal subspaces. In: Cryptographic Hardware and Embedded
  Systems - CHES 2006. vol.~4249, pp. 1--14. Springer, Heidelberg (October
  2006). \doi{10.1007/11894063\_1}

\bibitem{Armananzas2008review}
Armañanzas, R., Inza, I., Santana, R., Saeys, Y., Flores, J.L., Lozano, J.A.,
  Van~de Peer, Y., Blanco, R., Robles, V., Bielza, C., Larrañaga, P.: A review
  of estimation of distribution algorithms in bioinformatics. BIODATA MINING
  \textbf{1}, ~12 (2008), \url{http://dx.doi.org/10.1186/1756-0381-1-6}

\bibitem{Baluja1997UsingOD}
Baluja, S., Davies, S.: Using optimal dependency-trees for combinatorial
  optimization: Learning the structure of the search space (1997)

\bibitem{lejla2011mia}
Batina, L., Gierlichs, B., Prouff, E., Rivain, M., Standaert, F.X.,
  Veyrat-Charvillon, N.: {Mutual Information Analysis: a Comprehensive Study}.
  J. Cryptology  \textbf{24}(2),  269--291 (2011)

\bibitem{Bhasin2019MindTP}
Bhasin, S., Chattopadhyay, A., Heuser, A., Jap, D., Picek, S., Shrivastwa,
  R.R.: Mind the portability: A warriors guide through realistic profiled
  side-channel analysis. In: Network and Distributed System Security Symposium
  (January 2020). \doi{10.14722/ndss.2020.24390}

\bibitem{brier2004cpa}
Brier, E., Clavier, C., Olivier, F.: Correlation power analysis with a leakage
  model. In: Joye, M., Quisquater, J.J. (eds.) CHES 2004. pp. 16--29. Springer
  Berlin Heidelberg, Berlin, Heidelberg (2004)

\bibitem{Brumley2009cache}
Brumley, B.B., Hakala, R.M.: Cache-timing template attacks. In: Matsui, M.
  (ed.) Advances in Cryptology -- ASIACRYPT 2009. pp. 667--684. Springer Berlin
  Heidelberg, Berlin, Heidelberg (2009)

\bibitem{CNNagainstCM}
Cagli, E., Dumas, C., Prouff, E.: Convolutional neural networks with data
  augmentation against jitter-based countermeasures. In: Fischer, W., Homma, N.
  (eds.) Cryptographic Hardware and Embedded Systems - CHES 2017. pp. 45--68.
  Springer International Publishing (2017). \doi{10.1007/978-3-319-66787-4\_3}

\bibitem{cryptoeprint:2019:054}
Carbone, M., Conin, V., Cornelie, M.A., Dassance, F., Dufresne, G., Dumas, C.,
  Prouff, E., Venelli, A.: Deep learning to evaluate secure {RSA}
  implementations. Cryptology ePrint Archive, Report 2019/054 (2019)

\bibitem{Chari2002template}
Chari, S., Rao, J.R., Rohatgi, P.: Template attacks. In: Cryptographic Hardware
  and Embedded Systems - CHES 2002. pp. 13--28. Springer, Heidelberg (2002).
  \doi{10.1007/3-540-36400-5\_3}

\bibitem{Choundary2018efficient}
{Choudary}, M.O., {Kuhn}, M.G.: Efficient, portable template attacks. IEEE
  Transactions on Information Forensics and Security  \textbf{13}(2),  490--501
  (Feb 2018). \doi{10.1109/TIFS.2017.2757440}

\bibitem{TAonDD}
Choudary, O., Kuhn, M.G.: Template attacks on different devices. In:
  Constructive Side-Channel Analysis and Secure Design, COSADE, pp. 179--198.
  Springer International Publishing (2014). \doi{10.1007/978-3-319-10175-0\_13}

\bibitem{Coffin2008eda}
Coffin, D., Smith, R.E.: Linkage Learning in Estimation of Distribution
  Algorithms, pp. 141--156. Springer Berlin Heidelberg, Berlin, Heidelberg
  (2008). \doi{10.1007/978-3-540-85068-7_7},
  \url{https://doi.org/10.1007/978-3-540-85068-7_7}

\bibitem{CC}
{C}ommon {C}riteria: {C}ommon {C}riteria v3.1. {R}elease 5 (apr 2017),
  \url{https://www.commoncriteriaportal.org/cc/index.cfm?}, last accessed:
  December 31, 2020

\bibitem{daemen02}
Daemen, J., Rijmen, V.: {The Design of Rijndael: AES - The Advanced Encryption
  Standard (Information Security and Cryptography)}. Springer, 1 edn. (2002)

\bibitem{DeBonet1996many}
De~Bonet, J.S., Isbell, C.L., Viola, P.: Mimic: Finding optima by estimating
  probability densities. In: Proceedings of the 9th International Conference on
  Neural Information Processing Systems. p. 424–430. NIPS'96, MIT Press,
  Cambridge, MA, USA (1996)

\bibitem{Elaabid2012}
Elaabid, M. Abdelazizand~Guilley, S.: Portability of templates. Journal of
  Cryptographic Engineering  \textbf{2}(1),  63--74 (May 2012).
  \doi{10.1007/s13389-012-0030-6}

\bibitem{EMVCo}
EMVCo: {EMV} specifications (2001), \url{https://www.emvco.com/}, last
  accessed: December 31, 2020

\bibitem{LDA2}
Fisher, R.: The statistical utilization of multiple measurements. Annals of
  Eugenics (Cambridge)  \textbf{8},  376--386 (November 1935).
  \doi{10.1111/j.1469-1809.1938.tb02189.x}

\bibitem{Gierlichs2007mutual}
Gierlichs, B., Batina, L., Tuyls, P., Preneel, B.: {Mutual information
  analysis}. In: CHES. pp. 426--442. Springer (2008)

\bibitem{TAvsSTO}
Gierlichs, B., Lemke-Rust, K., Paar, C.: Templates vs. stochastic methods. In:
  Cryptographic Hardware and Embedded Systems - CHES 2006. Springer, Heidelberg
  (October 2006). \doi{10.1007/11894063\_2}

\bibitem{Gohr2019CHES2S}
Gohr, A., Jacob, S., Schindler, W.: {CHES} 2018 side channel contest {CTF} -
  solution of the {AES} challenges. IACR Cryptology ePrint Archive  (2019)

\bibitem{Gonzalez2002mathematical}
González, C., Lozano, J., Larranaga, P.: Mathematical modelling of umdac
  algorithm with tournament selection. behaviour on linear and quadratic
  functions. International Journal of Approximate Reasoning  \textbf{31},
  313--340 (11 2002). \doi{10.1016/S0888-613X(02)00092-0}

\bibitem{Gilbert2011tvla}
Goodwill, G., Jun, B., Jaffe, J., Rohatgi, P.: A testing methodology for
  side-channel resistance validation (2011)

\bibitem{Gruss2015cacheTA}
Gruss, D., Spreitzer, R., Mangard, S.: Cache template attacks: Automating
  attacks on inclusive last-level caches. pp. 897--912 (08 2015)

\bibitem{SVM2}
Heuser, A., Zohner, M.: Intelligent machine homicide. In: Constructive
  Side-Channel Analysis and Secure Design. pp. 249--264. Springer, Heidelberg
  (2012). \doi{10.1007/978-3-642-29912-4\_18}

\bibitem{SVM1}
Hospodar, G., Gierlichs, B., De~Mulder, E., Verbauwhede, I., Vandewalle, J.:
  Machine learning in side-channel analysis: a first study. J. Cryptographic
  Engineering  \textbf{1},  293--302 (Oct 2011).
  \doi{10.1007/s13389-011-0023-x}

\bibitem{LDA1}
Johnson, R.A., Wichern, D.W. (eds.): Applied Multivariate Statistical Analysis.
  Prentice-Hall, Inc., Upper Saddle River, NJ, USA (1988)

\bibitem{kim2018noise}
Kim, J., Picek, S., Heuser, A., Bhasin, S., Hanjalic, A.: Make some noise.
  unleashing the power of convolutional neural networks for profiled
  side-channel analysis. IACR Transactions on Cryptographic Hardware and
  Embedded Systems  \textbf{2019}(3),  148--179 (May 2019),
  \url{https://tches.iacr.org/index.php/TCHES/article/view/8292}

\bibitem{kocher1999dpa}
Kocher, P., Jaffe, J., Jun, B.: Differential power analysis. In: Wiener, M.
  (ed.) Advances in Cryptology --- CRYPTO' 99. pp. 388--397. Springer Berlin
  Heidelberg, Berlin, Heidelberg (1999)

\bibitem{keyboardAES}
Kwonyoup~Kim, Tae Hyun~Kim, T.K., Ryu, S.: {AES} wireless keyboard: Template
  attack for eavesdropping. In: Black Hat Asia, Singapore (2018)

\bibitem{Larranaga2002edas}
Larranaga, P., Lozano, J.: Estimation of Distribution Algorithms: A New Tool
  for Evolutionary Computation, Genetic algorithms and evolutionary
  computation, vol.~2. Springer US, 1st edn. (Jan 2002).
  \doi{10.1007/978-1-4615-1539-5}

\bibitem{SVM3}
Lerman, L., Bontempi, G., Markowitch, O.: Side channel attack : an approach
  based on machine learning. In: Constructive Side-Channel Analysis and Secure
  Design, COSADE (2011)

\bibitem{RF}
Lerman, L., Bontempi, G., Markowitch, O.: A machine learning approach against a
  masked aes. Journal of Cryptographic Engineering  \textbf{5}(2),  123--139
  (Jun 2015). \doi{10.1007/s13389-014-0089-3}

\bibitem{TAvsML}
Lerman, L., Poussier, R., Markowitch, O., Standaert, F.X.: Template attacks
  versus machine learning revisited and the curse of dimensionality in
  side-channel analysis: extended version. Journal of Cryptographic Engineering
   \textbf{8}(4),  301--313 (Nov 2018). \doi{10.1007/s13389-017-0162-9}

\bibitem{Lozano2006edas}
Lozano, J., Larranaga, P., Inza, I., Bengoetxea, E.: Towards a New Evolutionary
  Computation: Advances in the Estimation of Distribution Algorithms, Studies
  in Fuzziness and Soft Computing, vol.~192. Springer-Verlag Berlin Heidelberg,
  1st edn. (Jan 2006). \doi{10.1007/3-540-32494-1}

\bibitem{Maghrebi2016BreakingCI}
Maghrebi, H., Portigliatti, T., Prouff, E.: Breaking cryptographic
  implementations using deep learning techniques. In: SPACE 2016. pp. 3--26
  (December 2016). \doi{10.1007/978-3-319-49445-6\_1}

\bibitem{BlueBook}
Mangard, S., Oswald, E., Popp, T.: Power Analysis Attacks: Revealing the
  Secrets of Smart Cards. Springer (2007)

\bibitem{masure2019comprehensive}
Masure, L., Dumas, C., Prouff, E.: A comprehensive study of deep learning for
  side-channel analysis. Transactions on Cryptographic Hardware and Embedded
  Systems  \textbf{2020} (11 2019)

\bibitem{Muhlenbein1996edas}
M{\"u}hlenbein, H., Paa{\ss}, G.: From recombination of genes to the estimation
  of distributions i. binary parameters. In: Voigt, H.M., Ebeling, W.,
  Rechenberg, I., Schwefel, H.P. (eds.) Parallel Problem Solving from Nature
  --- PPSN IV. pp. 178--187. Springer Berlin Heidelberg, Berlin, Heidelberg
  (1996)

\bibitem{Muhlenbein1997equation}
Mühlenbein, H.: The equation for response to selection and its use for
  prediction. Evolutionary Computation  \textbf{5}(3),  303--346 (1997).
  \doi{10.1162/evco.1997.5.3.303},
  \url{https://doi.org/10.1162/evco.1997.5.3.303}

\bibitem{Oswald2007template}
Oswald, E., Mangard, S.: Template attacks on masking—resistance is futile.
  vol.~4377, pp. 243--256 (02 2007). \doi{10.1007/11967668_16}

\bibitem{Paguada2020controlling}
Paguada, S., Rioja, U., Armendariz, I.: Controlling the deep learning-based
  side-channel analysis: A way to leverage from heuristics. In: Applied
  Cryptography and Network Security Workshops. pp. 106--125. Springer
  International Publishing, Cham (2020)

\bibitem{Pelikan2002survey}
Pelikan, M., Goldberg, D., Lobo, F.: A survey of optimization by building and
  using probabilistic models. Computational Optimization and Applications
  \textbf{21},  5--20 (01 2002). \doi{10.1023/A:1013500812258}

\bibitem{Pelikan99boa}
Pelikan, M., Goldberg, D.E.: Boa: The bayesian optimization algorithm. In: in
  Proc. Genetic and. pp. 525--532 (1999)

\bibitem{Pelikan2003hierarchical}
Pelikan, M., Goldberg, D., Tsutsui, S.: Hierarchical bayesian optimization
  algorithm: toward a new generation of evolutionary algorithms. pp. 2738 --
  2743 Vol.3 (09 2003). \doi{10.1109/SICE.2003.1323811}

\bibitem{Pelikan1998bivariate}
Pelikan, M., Mühlenbein, H., Informationstechnik, G.: The bivariate marginal
  distribution algorithm  (07 1998). \doi{10.1007/978-1-4471-0819-1\_39}

\bibitem{Picek2019feature}
Picek, S., Heuser, A., Jovic, A., Batina, L.: A systematic evaluation of
  profiling through focused feature selection. IEEE Transactions on Very Large
  Scale Integration (VLSI) Systems  \textbf{PP},  1--14 (09 2019).
  \doi{10.1109/TVLSI.2019.2937365}

\bibitem{picek2018ontheperformance}
Picek, S., Samiotis, I.P., Kim, J., Heuser, A., Bhasin, S., Legay, A.: On the
  performance of convolutional neural networks for side-channel analysis. In:
  Chattopadhyay, A., Rebeiro, C., Yarom, Y. (eds.) Security, Privacy, and
  Applied Cryptography Engineering. pp. 157--176. Springer International
  Publishing, Cham (2018)

\bibitem{CNNOV}
Picek, S., Samiotis, I.P., Kim, J., Heuser, A., Bhasin, S., Legay, A.: On the
  performance of convolutional neural networks for side-channel analysis. In:
  Security, Privacy, and Applied Cryptography Engineering. pp. 157--176.
  Springer International Publishing, Cham (2018)

\bibitem{prouff2018ascad}
Prouff, E., Strullu, R., Benadjila, R., Cagli, E., Canovas, C.: Study of deep
  learning techniques for side-channel analysis and introduction to ascad
  database. IACR Cryptol. ePrint Arch.  \textbf{2018}, ~53 (2018)

\bibitem{Rechberger2005PracticalTA}
Rechberger, C., Oswald, E.: Practical template attacks. In: Lim, C.H., Yung, M.
  (eds.) Information Security Applications. pp. 440--456. Springer, Heidelberg
  (2005). \doi{10.1007/978-3-540-31815-6\_35}

\bibitem{Rechberger2005Practical}
Rechberger, C., Oswald, E.: {Practical Template Attacks}. In: Information
  Security Applications. Springer (Jan 2005). \doi{DOItmp_0558_026432},
  \url{http://dx.doi.org/DOItmp_0558_026432}

\bibitem{FormalStudyPowerVariability}
Renauld, M., Standaert, F.X., Veyrat-Charvillon, N., Kamel, D., Flandre, D.: A
  formal study of power variability issues and side-channel attacks for
  nanoscale devices. In: Advances in Cryptology -- EUROCRYPT 2011. pp.
  109--128. Springer, Heidelberg (2011). \doi{10.1007/978-3-642-20465-4\_8}

\bibitem{Rioja2020similarity}
Rioja, U., Batina, L., Armendariz, I.: When similarities among devices are
  taken for granted: Another look at portability. In: Nitaj, A., Youssef, A.
  (eds.) Progress in Cryptology - AFRICACRYPT 2020. pp. 337--357. Springer
  International Publishing, Cham (2020)

\bibitem{ePrintSixSigma}
Rioja, U., Paguada, S., Batina, L., Armendariz, I.: The uncertainty of
  side-channel analysis: A way to leverage from heuristics. {IACR} Cryptology
  ePrint Archive

\bibitem{Santana1999edge}
Santana~R, Ponce de León~E, O.A.: The edge incident model. In: Proceedings of
  the Second Symposium on Artificial Intelligence (CIMAF-99). pp. 352--359
  (1999)

\bibitem{stochastic}
Schindler, W., Lemke, K., Paar, C.: A stochastic model for differential side
  channel cryptanalysis. In: Cryptographic Hardware and Embedded Systems - CHES
  2005. pp. 30--46. Springer, Heidelberg (2005). \doi{10.1007/11545262\_3}

\bibitem{Schwarz2019JavaScriptTA}
Schwarz, M., Lackner, F., Gruss, D.: Javascript template attacks: Automatically
  inferring host information for targeted exploits. In: NDSS (2019)

\bibitem{Standaert2008Using}
Standaert, F.X., Archambeau, C.: Using subspace-based template attacks to
  compare and combine power and electromagnetic information leakages. In:
  Oswald, E., Rohatgi, P. (eds.) Cryptographic Hardware and Embedded Systems -
  CHES 2008. pp. 411--425. Springer, Heidelberg (2008).
  \doi{10.1007/978-3-540-85053-3\_26}

\bibitem{Standaert2009UnifiedFramework}
Standaert, F.X., Malkin, T.G., Yung, M.: {A Unified Framework for the Analysis
  of Side-Channel Key Recovery Attacks}. In: Joux, A. (ed.) Advances in
  Cryptology - EUROCRYPT 2009. pp. 443--461. Springer Berlin Heidelberg,
  Berlin, Heidelberg (2009)

\bibitem{UCdatasheet}
STMicroelectronics: Stm32f411vet6 datasheet (Dec 2016),
  \url{https://www.alldatasheet.com/datasheet-pdf/pdf/929991/STMICROELECTRONICS/STM32F411VET6.html},
  last accessed: November 11, 2020

\bibitem{Tanco2009doe}
Tanco, M., Viles, E., Ilzarbe, L., Alvarez, M.: Implementation of design of
  experiments projects in industry. Applied Stochastic Models in Business and
  Industry  \textbf{25},  478 -- 505 (07 2009)

\bibitem{zaid2019methodology}
Zaid, G., Bossuet, L., Habrard, A., Venelli, A.: {Methodology for Efficient CNN
  Architectures in Profiling Attacks}. IACR Transactions on Cryptographic
  Hardware and Embedded Systems  \textbf{Volume 2020},  Issue 1-- (2019)

\end{thebibliography}

\end{document}